\documentclass{amsproc}
\usepackage{amscd}
\usepackage{graphics}
\newtheorem{Theorem}{Theorem}[section]
\newtheorem{Proposition}{Proposition}[section]

\newtheorem{Lemma}{Lemma}[section]
\newtheorem{conjecture}[Theorem]{Conjecture}
\def\proof{\par{\it Proof}. \ignorespaces}
\def\endproof{{\ \vbox{\hrule\hbox{%
   \vrule height1.3ex\hskip0.8ex\vrule}\hrule }}\par}
\newenvironment{Proof}{\proof}{\endproof}

\theoremstyle{definition}
\newtheorem{Definition}[Theorem]{Definition}
\newtheorem{Example}[Theorem]{Example}
\newtheorem{Notation}[Theorem]{Notation}

\theoremstyle{remark}
\newtheorem{remark}[Theorem]{Remark}

\numberwithin{equation}{section}

\begin{document}

\title{Blow-ups of the Toda lattices and their intersections with the
Bruhat cells}

\author{Luis Casian}
\address{Department of Mathematics, Ohio State University, Columbus,
OH 43210}
\email{casian@math.ohio-state.edu}
\thanks{Both authors are supported in part by NSF Grant \#DMS0071523.}

\author{Yuji Kodama}
\address{Department of Mathematics, Ohio State University, Columbus,
OH 43210}
\email{kodama@math.ohio-state.edu}

\date{January 22, 2002 and, in revised form, }


\begin{abstract}
We study the topology of the set of singular points (blow-ups) in the
solution of the nonperiodic Toda lattice defined on real split semisimple
Lie algebra $\mathfrak g$.
The set of blow-ups is called the Painlev\'e divisor. The isospectral
manifold of the
Toda lattice is compactified through the companion embedding which maps the
manifold
to the flag manifold associated with the underlying Lie algebra $\mathfrak g$.
The Painlev\'e divisor is then given by the intersections of the compactified
manifold with the Bruhat cells in the flag manifold. In this paper, we give
explicit
description of the topology of the Painlev\'e divisor for the cases of all
the rank two
Lie algebra, $A_2,B_2, C_2, G_2$, and $A_3$ type. The results are obtained by
using the Mumford system and the limit matrices introduced originally for
the periodic
Toda lattice. We also give a Lie theoretic description
of the Painlev\'e divisor of codimension one case, and propose several
conjectures
for the general case.

\end{abstract}

\maketitle
\markboth{LUIS CASIAN AND YUJI KODAMA}
  {BLOW-UPS OF THE TODA LATTICES}
\tableofcontents

\section{Introduction}
It is well-known that the generalized (nonperiodic) Toda lattices asociated
with semisimple Lie algebra $\mathfrak g$ of rank $l$ possess $l$ polynomial
invariants, {\it the Chevalley invariants}, which provide their
integrability
\cite{bogoyavlensky:76, kostant:79}. The isospectral manifold determined by
those
polynomials defines a $l$-dimensional affine variety, and it can be
compactified by adding the points associated with the {\it blow-ups} in the
solution of the Toda lattice. Those points are defined as the zeros of
$\tau$-functions giving an explicit solution of
the Toda lattice \cite{jimbo:83}, and the set of zeros is sometimes called
the {\it Painlev\'e divisor}. The number of $\tau$-functions is given by
the rank of the algebra, and each $\tau$-function can be labeled by a dot in
the corresponding Dynkin diagram. Then the Painlev\'e divisor consists of
$l$ components
$\{\Theta_{\{k\}}:k=1,\cdots,l\}$, and each $\Theta_{\{k\}}$ is associated
with a root $\alpha_k$ in the Dynkin diagram. As in the case of periodic
Toda \cite{adler:91}, the singularities of the divisor are canonically
associated with the Dynkin diagrams, i.e.
$\Theta_J=\cap_{k\in J}\Theta_{\{k\}}$ for a subdiagram $J\subset
\{1,\cdots,l\}$.

In \cite{flaschka:91}, Flaschka and Haine considered a {\it companion} embedding
map of the isospectral manifold into a flag manifold, and identified
the Painlev\'e divisor as the intersection with certain Bruhat cells in the
Bruhat decomposition,
\[ G/B^+=\bigcup_{w\in W} N^-w B^+/B^+,\]
where $G$ is the Lie group with ${\mathfrak g}={\rm Lie}(G)$, $B^+$ the
Borel subgroup, $N^-$ the unipotent subgroup and $W$ the Weyl group of $G$.
Then the compactification of the isospectral manifold can be obtained by
gluing the Painlev\'e divisor in the flag manifold.

On the other hand, the real part of the compactified isospectral manifold
was studied in \cite{casian:99}, where  the manifold was constructed by
extending the work of Kostant in \cite{kostant:79}.  Theorem 3.2 in
\cite{kostant:79} describes part of the  isospectral manifold of the Toda
lattice
in terms of one connected component of a split Cartan subgroup of $G$.
There is a total of $2^l$ connected components which are labeled by a set of
signs
$\epsilon=(\epsilon_1,\cdots,\epsilon_l)$. In \cite{casian:99} instead all
the connected components of a
Cartan subgroup  are involved. The upshot is that now a split Cartan
subgroup $H_{\mathbb R}$, with all its connected components, becomes an open
dense
subset in the compactified isospectral manifold. This manifold  is then
described as a union of convex polytopes $\Gamma_\epsilon$ glued as
in \cite{twisted}, and each connected component with the sign $\epsilon$ of
the Cartan subgroup is the interior of the corresponding polytope
$\Gamma_{\epsilon}$. The convexity of the polytope $\Gamma_{\epsilon}$
can be shown by Atiyah's convexity theorem \cite{atiyah:82} with the torus
embedding (conjugate to the companion embedding) in the flag manifold.

In this paper, we study the topological structure of the Painlev\'e divisor
as the blow-ups
of the Toda lattice on the polytopes. In Section \ref{chev}, we provide a
background
information necessary for the present study which includes the isospectral
manifold, the companion embedding to the flag manifold, the $\tau$-functions
and the Painlev\'e divisor.

In Section \ref{painlevediv}, we define the {\it limit matrices} to parametrize
the Painlev\'e divisor. The limit matrix was first introduced in
\cite{adler:93} for the periodic Toda lattice for a parametrization of the
Birkhoff strata of the hyperelliptic Jacobi variety,
and the existence of the limit matrix was shown based on Sato's theory of
universal Grassmannians.
We here give a direct proof of the existence of the limit matrix by using a
factorization of the unipotent subgroup $N^-$ (Proposition
\ref{limitmatrixAD}), and show that the companion embedding maps the limit
matrix to the corresponding Bruhat cell.

In Section \ref{mumfordsect}, we define the Mumford system for the $A_l$
Toda lattice,
which may be considered as an extension of the system used to parametrize the
moduli
space associated with the hyperelliptic Riemann surface and its Jacobian.
The Mumford system gives an explicit coordinate for the Painlev\'e divisor
through the limit matrix.
Then we prove a topological equivalence between the top cell of $A_k$
and certain Painlev\'e divisor of $A_j$ with $j>k$ (Proposition
\ref{divisorequivalenceD}).

In Section \ref{rank2example}, we provide several explicit results for the
Toda lattices
on the Lie algebra $\mathfrak g$ of all rank 2 cases, $A_2, B_2, C_2, G_2$,
and
of type $A_3$.

Then in Section \ref{negativesection}, we give a Lie
theoretic description of the Painlev\'e divisor based on the results in
\cite{casian:99, twisted}.
We first review the details of the
construction of the compactified manifold by gluing the polytopes
$\Gamma_{\epsilon}$ of the Cartan subgroup $H_{\mathbb R}$,  and it is
worth keeping in mind that $H_{\mathbb R}$  is not necessarily a Cartan
subgroup in $G$ but rather in another Lie group
$\tilde G$ defined in Notation  \ref{standardG}.
We then define an \lq\lq algebraic\rq\rq
version of the
Painlev\'e divisor, denoted by ${\tilde\Theta}_{\{i\}}^a$, in terms of the
simple root character $\chi_{\alpha_i}$ defined on the Cartan subgroup. The
characters $\chi_{\alpha_k}$ can be expressed in terms of the characters
$\chi_{\omega_i}$ associated to the fundamental weights $\omega_i$ which
have similar properties to the $\tau$-functions. Then we give Conjectures
\ref{conjecture1} and \ref{conjecture2} that $\Theta_{ \{ i \} }$ and
$\tilde \Theta^a_{ \{ i \} } $ become homeomorphic if  small modifications
on $\tilde \Theta^a_{ \{ i \} } $ are introduced.
These conjectures about the structure of the Painlev\'e divisors are
 verified in all the rank 2 cases as well as in $A_3$ discussed in Section
\ref{rank2example}. The  homology of the  spaces
 constructed in terms of the root characters is computable with the same
methods used in \cite{casian:99}. Conjectures
 \ref{conjecture1} and \ref{conjecture2}  then would allow the computation
of the homology of the Painlev\'e divisors.

\section{Toda lattices and Painlev\'e divisor \label{chev}}
The generalized (nonperiodic) Toda lattice equation related to
real split semisimple Lie algebra $\mathfrak g$ of rank $l$ is defined by
the Lax equation,
\cite{bogoyavlensky:76,kostant:79},
\begin{equation}
\label{lax}
\displaystyle{{dL \over dt}=[A, L]}
\end{equation}
where the Lax pair $(L,A)$ are given by
\begin{equation}
\left\{
\begin{array} {ll}
& \displaystyle{L(t)=\sum_{i=1}^l b_i(t)h_{\alpha_i}+\sum_{i=1}^l \left(
a_i(t)e_{-\alpha_i}+e_{\alpha_i}\right)} \\
& \displaystyle{A(t)=-\sum_{i=1}^l a_i(t)e_{-\alpha_i}}
\end{array}
\right.
\label{LA}
\end{equation}
Here $\{h_{\alpha_i},e_{\pm\alpha_i}\}$ is the Cartan-Chevalley basis of the
algebra $\mathfrak g$ with the positive simple roots
$\Pi=\{\alpha_1,\cdots,\alpha_l\}$
which satisfy the relations,
\[
  [h_{\alpha_i} , h_{\alpha_j}] = 0, \quad
  [h_{\alpha_i}, e_{\pm \alpha_j}] = \pm C_{j,i}e_{\pm \alpha_j} \ , \quad
  [e_{\alpha_i} , e_{-\alpha_j}] = \delta_{i,j}h_{\alpha_j},
\]
where $(C_{i,j})$ is the $l\times l$ Cartan matrix of the Lie algebra
$\mathfrak g$.
The Lax equation (\ref{lax}) then gives
\begin{equation}
\left\{
\begin{array} {ll}
& \displaystyle{{d b_i \over dt}=a_i} \\
& \displaystyle{{d a_i \over dt}=  -\left(\sum_{j=1}^l C_{i,j} b_j\right)a_i}
\end{array}
\right.
\label{toda-lax}
\end{equation}

The integrability of the system can be shown by the existence of the Chevalley
invariants, $\{I_k(L):k=1,\cdots,l\}$, which are given by the homogeneous
polynomial
of $\{(a_i, b_i): i=1,\cdots,l\}$. Then in this paper we are concerned with the
topology of the {\it real} isospectral manifold defined by
\[
Z(\gamma)_{\mathbb R}=\left\{(a_1,\cdots,a_l,b_1\cdots,b_l)\in{\mathbb
R}^{2l}~:~
I_k(L)=\gamma_k\in {\mathbb R},~ k=1,\cdots,l\right\}.
\]
The manifold $Z(\gamma)_{\mathbb R}$ can be compactified by adding the set of
points corresponding to the {\it blow-ups} of the solution.
The set of blow-ups has been shown to be characterized by the intersections
with the Bruhat cells of
the flag manifold $G/B^+$, which are referred to as the
{\it Painlev\'e divisors}, and the compactification is described in the
flag manifold. In order to explain some details of this fact,
we first define the set ${\mathcal F}_{\gamma}$,
\[
{\mathcal F}_{\gamma}:= \{ L\in e_++{\mathcal B}^-~:~ I_k(L)=\gamma_k,
k=1,\cdots, l\},
\]
where $e_+=\sum_{i=1}^le_{\alpha_i}$, and ${\mathcal B}^- $ is the Lie
algebra of $B^-$. Then there exists a unique element $n_0\in N^-$ such that
$L\in{\mathcal F}_{\gamma}$ can be conjugated to the normal form $C_{\gamma}$,
$L=n_0C_{\gamma}n_0^{-1}$ \cite{kostant:78}. In the case of ${\mathfrak g}=
{\mathfrak sl}(l+1,{\mathbb R})$, $C_{\gamma}$ is the companion matrix given by
\[
C_{\gamma}=\left(
\begin{matrix}
0 &  1 &  0 &  \cdots &  0 \\
0 & 0 &  1 &  \cdots & 0 \\
\vdots &  \ddots & \ddots &  \ddots &  \vdots \\
0 &  \cdots & \ddots & 0 & \ 1 \\
(-1)^{l}{\gamma}_l&  \cdots & \cdots & -{\gamma}_1 & 0 \\
\end{matrix}
\right),
\]
where the Chevalley invariants are given by the elementary
symmetric polynomials of the eigenvalues of $L$. Then we define:
\begin{Definition}\cite{flaschka:91}:
The companion embedding of ${\mathcal F}_{\gamma}$ is defined as the map,
\label{Cembedding}
\[
\begin{matrix}
c_{\gamma}: &{\mathcal F}_{\gamma} & \longrightarrow & G/B^+ \\
{}          & L  & \longmapsto & n_0^{-1} ~{\rm mod} B^+
\end{matrix}
\]
where $ L=n_0C_{\gamma}n_0^{-1}$ with $n_0\in N^-$.
\end{Definition}
The isospectral manifold $Z(\gamma)_{\mathbb R}$ can be considered as a subset
of ${\mathcal F}_{\gamma}$ with the element $L$ in the form of (\ref{LA}).
The Toda lattice (\ref{lax}) then defines a flow on ${\mathcal F}_{\gamma}$
which is embedded
as follows:
\begin{Proposition}\cite{flaschka:91}
\label{todainflag}
The Toda flow maps to the flag manifold as
\[
\begin{matrix}
c_{\gamma} : &L(t) & \longmapsto & n_0^{-1}n(t)~ &~ {\rm mod} ~B^+ \\
 {}          & {}  &    {}       & = n_0^{-1} e^{tL^0} ~&~{\rm mod}~B^+
 \end{matrix}
\]
where $L^0=n_0C_{\gamma}n_0^{-1}$, and $n(t)\in N^-, b(t)\in B^+$ are given
by the
factorization of $e^{tL^0}=n(t)b(t)$.
\end{Proposition}
This Proposition is based on the solution formula using the factorization, i.e.
\begin{equation}
\label{solution}
L(t)=n(t)^{-1}L^0n(t)=b(t)L^0b(t)^{-1}.
\end{equation}
However one should note
that the factorization is not always possible,
and the general form is given by the Bruhat decomposition,
\[
G=\bigcup_{w\in W}N^-wB^+\ .
\]
It has been also shown in
\cite{flaschka:91, adler:91}
that for a subset $J$ of $\{1,\cdots,l\}$ the
blow-up of the solution $L(t)$ at $t=t_J$ corresponds to the case
\[
e^{t_J L^0} \in N^-w_J B^+, \quad {\mbox {where}} \ \ w_J \neq id \ ,
\]
where $w_J$ is the longest element of the Weyl subgroup $W_J$ associated
with the Dynkin diagram labeled by $J$. Thus the Toda flow meets only those
Bruhat cells, and we see
that the Painlev\'e divisor, denoted by ${\mathcal D}_J$, characterizes the
intersection of the Bruhat cell
corresponding to the longest element $w_J\in W$ with the compactified
isospectral manifold ${\tilde Z}(\gamma)_{\mathbb R}$, i.e.
\begin{equation}
\label{bruhatcell}
{\mathcal D}_J={\tilde Z}(\gamma)_{\mathbb R}\bigcap N^-w_J B^+/B^+, \quad {\rm
with} \quad w_{\emptyset}=id.
\end{equation}
Here ${\tilde Z}(\gamma)_{\mathbb R}$ is the closure of the image of the
isospectral manifold under the companion embedding $c_{\gamma}$ in
(\ref{Cembedding}), and it has a decomposition (intersection with the
Bruhat decomposition),
\[
{\tilde Z}(\gamma)_{\mathbb R}=\overline{c_{\gamma}(Z(\gamma)_{\mathbb R})}
=\bigsqcup_{J\subset\{1,\cdots,l\}}{\mathcal D}_J.
\]

The analytical structure of the blow-ups can be obtained by the
$\tau$-functions,
which are defined by
\begin{equation}
\label{tau}
b_k=\displaystyle{{d\over dt}\ln \tau_k, \quad \quad a_k=a_k^0\prod_{j=1}^l
(\tau_j)^{-C_{k,j}}},
\end{equation}
 From (\ref{toda-lax}), the tau-functions then satisfy the bilinear equations,
\begin{equation}
\label{bilinear}
\tau_k\tau_k''-(\tau_k')^2=\prod_{j\ne k}(\tau_j)^{-C_{k,j}},
\end{equation}
where $\tau_k''=d^2\tau_k/dt^2$ and $\tau_k'=d\tau_k/dt$, and $\tau_0=1,
\tau_{l+2}=0$.
Then the Painlev\'e divisor ${\mathcal D}_J$ can be defined as
\begin{equation}
\label{painlevedivisor}
c_{\gamma}(L(t))\in {\mathcal D}_J
\overset{\rm def}{\iff} \tau_k(t)=0, ~{\rm iff} ~k\in J.
\end{equation}
We also define the set $\Theta_J$ as a disjoint union of ${\mathcal D}_{J'}$,
\[
\Theta_J:=\bigsqcup_{J'\supseteq J}{\mathcal D}_{J'} .\]
Then we have a stratification of ${\tilde Z}(\gamma)_{\mathbb R}$,
\[
{\tilde Z}(\gamma)_{\mathbb R}=\Theta^{(0)}\supset \Theta^{(1)}\supset \cdots
\supset \Theta^{(l)}=c_{\gamma}(C_{\gamma}), \quad {\rm with} \quad
\Theta^{(k)}=\bigcup_{|J|=k}\Theta_J. \]
The irreducibility of the Painlev\'e divisors $\Theta_{\{k\}}$ was shown
in \cite{ercolani:92}, where the analog of Riemann's singularity theorem
for the compactified complex manifold ${\tilde Z}(\gamma)_{\mathbb C}$
was also discussed.

In the case of a given matrix (adjoint) representation, one can construct
an explicit solution for $\{a_j(t)\}$ in the matrix $L(t)$. First we have the
following Lemma:
\begin{Lemma}
\label{bj}
The diagonal element $b_{j,j}$ of the upper triangular
matrix $b\in B^+$ in the factorization (\ref{todainflag})
is expressed by
\[
b_{j,j}(t)={D_j[\exp(tL^0)] \over D_{j-1}[\exp(tL^0)]}
\]
where $D_j[\exp(tL^0)]$ is the determinant of the $j$-th principal minor of
$\exp(tL^0)$, i.e.
\[
D_j[\exp(tL^0)]=\left( e^{tL^0}v_1\wedge\cdots\wedge v_j,\
v_1\wedge\cdots\wedge v_j\right).
\]
with the standard basis $\{v_i\}_{i=1}^l$ of ${\mathbb R}^n$ with some $n$.
\end{Lemma}
Then using the formula in (\ref{solution}), we can obtain the solution
$a_j(t)$ and the explicit representation of the
$\tau$-functions in terms of the determinants
$D_j[\exp(tL^0)]$. Thus the $\tau$-functions are the entire functions
of $t$ given by polynomials of exponential functions $\exp(\lambda_k t)$
with the eigenvalues $\lambda_k$ of $L^0$. In fact, one can show that the
$D_j[\exp(tL^0)]$
can be expressed as the Hankel determinant,
\[
D_j[\exp(tL^0)]=\left|\begin{matrix}
D_1 & D_1' & \cdots & D_1^{(j-1)}\\
D_1'& D_1''& \cdots & D_1^{(j)}\\
\vdots& \ddots & \ddots& \vdots \\
D_1^{(j-1)}& \cdots& \cdots& D_1^{(2j-2)}
\end{matrix}\right|, \quad j=1,2,\cdots, n, \]
where $D_1=D_1[\exp(tL^0)]=\sum_{i=1}^{n} \rho_i\exp(\lambda_i t)$
for some $\rho_i\in {\mathbb R}\setminus \{0\}$. With
this formula,
one can study a detailed behavior of the $\tau$ functions \cite{kodama:98}.

\begin{remark} On  any Cartan subgroup of $G$ there is another set of
functions having similar properties to the $\tau$ functions. These are the
root
characters $\chi_{\omega_i}$ associated to fundamental weights ${\omega_i}$.
For example, the simple root characters
$\chi_i:=\chi_{\alpha_i}$ can be expressed in terms of the $\chi_{\omega_i}$
with the inverse of the Cartan matrix
of the Lie algebra ${\mathfrak g}$.  This is the same relation that exists
between the $a_i$ in (\ref{LA})
and the $\tau$ functions. The signs of the characters $\chi_i$ change when
chamber walls $\chi_i=-1$  are crossed in a Cartan subgroup
in analogy to what happens to the signs of the $a_i$ when a Painlev\'e
divisor is crossed. If $\chi^*_i$ denotes the root character 
of the simple root $\alpha_i$ corresponding to each separate chamber
in the Cartan subgroup, 
then $\chi^*_i$ is continuous through $\alpha_i$ walls  and through some
$\alpha_j$ walls. The points on a Cartan subgroup where $\chi^*_i+1=0$ are
called the $\alpha_i -${\em negative wall} \cite{casian:99}, which
defines an \lq\lq algebraic\rq\rq version of the Painlev\'e divisors
 $\Theta_{ \{ i \} }$ in terms of the functions $\chi_i$.  This set is
compactified and gives rise to a topological
space $\tilde \Theta^a_{ \{ i \} } $ (see Section \ref{negativesection}).
\end{remark}

\section{Limit matrices, Painlev\'e divisors  and the companion embedding
\label{painlevediv}}
Here we show that Painlev\'e divisors  can be
parametrized using limit matrices. These were first introduced in
\cite{adler:93}
for the case of the  periodic Toda lattice.
The main result in \cite{adler:93} is to show the existence of the limit
matrix, say $L_J$, which is constructed by conjugating the Lax matrix $L(t)$
with a matrix in $N^-$ and taking the limit $t\to t_J$ corresponding to the
factorization $e^{t_JL^0}={\hat n}(t_J)w_J{\hat b}(t_J)$ for ${\hat n}\in N^-$
and ${\hat b}\in B^+$. In our case of the nonperiodic
Toda lattice  limit matrices arise as a consequence of
 Theorem 3.3 of \cite{flaschka:91}.

\begin{Definition}\label{piJ} For fixed $J\subset \{1,\cdots,l\}$ we let
$P_J$ denote the parabolic subgroup
of $ G$ containing $B^+$ and associated to $J$. One can define a projection
\[\pi_J: G/B^+ \to G/P_J.\]
The group $N^{-}$ factors as $N^{-}=N^{-}_JN^{+}_J$ with
$N^{\pm}_J:=N^{-}\cap w_J N^{\pm} w_J^{-1}$. Hence any $n\in N^-$
can be written as
$n=uy$ with $u\in N^{-}_J$ and $y\in N^{+}_J$ unique elements.
We thus obtain factorizations (notation of Proposition \ref{todainflag}):
$n_0^{-1}n(t)=u(t)y(t)$, and $\pi_J(u(t)y(t)B^+)=u(t)P_J$.
\end{Definition}

Since the limit $n^{-1}_0 n(t)B^{+}$ as $t\to t_J$ exists (see Proposition
\ref{todainflag}), it is of the form
 $\hat u(t_J) w_JB^{+}$ for some $\hat u(t_J)\in N^{-}_J$. Then we have

\begin{Proposition} With notation as in Definition \ref{piJ}, the limit of
$u(t)$ as $t\to t_J$ exists,
\[\lim_{t\to t_J} u(t)= \hat u(t_J)\in N^{-}_J.\]
\label{aboutU}
\end{Proposition}
\begin{proof}  Since $\lim_{t\to t_J} n^{-1}_0 n(t)B^{+}=\hat
u(t_J)w_JB^+$ with $\hat u(t_J)\in N^{-}_J$, by applying $\pi_J$
we obtain
$$\lim_{t\to t_J} n^{-1}_0n(t)P_J=\hat u(t_J)P_J.$$
On the other hand $\pi_J( n^{-1}_0n(t)B^{+})=\pi_J(u(t)y(t)B^{+})=u(t)P_J$.
Therefore, since the top $N^{-}$ orbit in $G/P_J$ can be identified with
$N^{-}_J$ we then obtain a limit inside this group:
 $\lim_{t\to t_J} u(t)= \hat u(t_J)$.
\end{proof}

\begin{Definition} A limit matrix of $L$ is an element $L_J$ in the set
${\mathcal F}_{\gamma}$ of the form,
\[L_J=Ad(\hat u^{-1}(t_J)) C_\gamma ,\quad {\rm with}\quad {\hat
u}(t_J)\in N^{-}_J.\]
\label{limitmatrix1}
\end{Definition}

 Let $u(t)=\hat u(t_J)\overline u(t)$. Then $\lim_{t\to t_J} \overline
u(t)=e$, with $e$ the identity, and we have

\begin{Proposition} The limit matrix is also expressed as
 \[\lim_{t\to t_J} Ad(y(t))L(t)=L_J(t_J).\]
\label{limitmatrixAD}
\end{Proposition}
\begin{proof} We have $$L(t)=Ad( n^{-1}(t) n_0)C_\gamma =
Ad(y^{-1}(t) u^{-1}(t) ) C_\gamma= Ad(y^{-1}(t)\overline u^{-1}(t)\hat
u^{-1}(t_J))C_\gamma$$
Hence $Ad(y(t))L(t)=Ad(\overline u(t))L_J$. We now take limit and use that
$\overline u(t) \to e$ to conclude.
\end{proof}
The result can be summarized in the diagram,
\[
\begin{CD}
L(t) @>{c_{\gamma}}>> n_0^{-1}n(t) ~{\rm mod} B^+\\
@V{}VV @V{}VV\\
Ad(y(t))L(t) @>>>  u(t)y(t)~{\rm mod} B^+\\
@V{t\to t_J}VV  @VV{t\to t_J}V\\
L_J(t_J) @>{c_{\gamma}}>> \hat u(t_J)w_J~{\rm mod} B^+
\end{CD}
\]

\begin{remark} For each set $J$ we can define a function $\phi_J:
Z(\gamma)_{\mathbb R}\to Ad( N^{-}_J) C_\gamma$ given
by  $\phi_J (L)=Ad(y)L$.
A limit matrix $L_J$ is then an element in the boundary
$\overline{\phi_J(Z(\gamma)_{\mathbb R}) } \setminus
\phi_J(Z(\gamma)_{\mathbb R})$.
The closure takes place inside $Ad( N^{-}_J) C_\gamma$. This gives another
description of ${\mathcal D}_J$
which allows one to define the compactification of the  isospectral manifold
$Z(\gamma)_{\mathbb R}$
using only the limit matrices. The companion embedding then takes a simple
form. First note that any limit matrix
$L_J$ is contained in the $ N^{-}_J$ orbit of  $C_\gamma.$
Hence $L_J=Ad({\hat u}^{-1}(t_J))C_\gamma$
where $\hat u(t_J) \in N^{-}_J$ is unique. For $J=\emptyset$, we just set
$\hat u(t)=n_0^{-1}n(t)$.
The companion embedding then maps $L_J$ to $\hat u(t_J) w_JB^+$.
\end{remark}

\begin{remark} In all our examples $y(t)=y_J(t)$ can be replaced with
$x_J^{-1}(t)$ an element in $N^{-}_J$
defined below in terms of a companion matrix associated to a Levi factor.
\end{remark}

In the following, we determine the limit matrices for the case of
${\mathfrak g}={\mathfrak sl}(l+1,{\mathbb R})$ (the general case will be
discussed elsewhere).
Let consider the set $J$ be given by $s$ consecutive
numbers, say $\{i+1,\cdots,i+s\},~(i+s\le l)$. Then from
(\ref{painlevedivisor})
this implies that the divisor ${\mathcal D}_J$ consists of the points
corresponding to the zeros of $\tau$-functions, $\tau_{k}=0$ for all $k\in
J$.
On the other hand, from (\ref{bilinear}), we can show
\begin{Lemma}
\label{multiplicity}
For each $j\in J=\{i+1,\cdots,i+s\}$,
$\tau_{j}(t)$ has the following form near its zero $t=t_J$,
\begin{equation}
\label{taukzero}
\tau_{i+k}(t)\simeq (t-t_J)^{m_{k}}+\cdots, \quad {\rm with}\quad m_k=k(s+1-k),
~ 1\le k\le s.
\end{equation}
\end{Lemma}
\begin{Proof}
Substituting (\ref{taukzero}) into (\ref{bilinear}), and using
$\tau_i(t_J)\ne 0$, we have $m_k=k(m_1+1-k)$.
Then $\tau_{i+s+1}(t_J)\ne 0$ implies $m_1=s$.
\end{Proof}
Then using (\ref{tau}) one can find the blow-up structure of the functions
$(a_j,b_j)$.
We note here that this structure is the same as the case of the smaller system
${\mathfrak sl}(s+1,{\mathbb R})$ with the total blow-up. The Lax matrix of
this smaller system
is just the submatrix (here the $b$-variables are modified from the original
form in
(\ref{LA}), e.g. $b_k-b_{k-1} \to b_k$),
\[L'=\left(
\begin{matrix}
b_{i+1} &  1      &  0     & \cdots     &  0 \\
a_{i+1} & b_{i+2} &  1     &  \cdots    &  0 \\
\vdots  &  \ddots & \ddots &  \ddots    &  \vdots \\
0       &  \cdots & \ddots & b_{i+s}    & \ 1 \\
0       &  \cdots & \cdots &   a_{i+s}  & b_{i+s+1} \\
\end{matrix}
\right)
\]
Then one can put this matrix into a companion matrix by a unique element
$x'_J\in N^-$,
the set of $(s+1)\times (s+1)$ lower triangular matrices with 1's on the
diagonals.
The companion matrix $C'_J={x'}^{-1}_JL'x'_J$ and $x'_J$ are given by
\[C'_J=\left(\begin{matrix}
0 &  1 &  0 &  \cdots &  0 \\
0 & 0 &  1 &  \cdots & 0 \\
\vdots &  \ddots & \ddots &  \ddots &  \vdots \\
0 &  \cdots & \ddots & 0 & \ 1 \\
(-1)^{s}{\xi}_{s+1}&  \cdots & \cdots & -{\xi}_2 & \xi_1 \\
\end{matrix}
\right), \quad x'_J=\left(\begin{matrix}
1      & 0     & \cdots &\cdots & 0 \\
*      & 1     & \ddots & \ddots& \vdots \\
\vdots & \ddots& \ddots & \ddots& \vdots \\
\vdots & \ddots& \ddots & 1     & 0    \\
*      & \cdots& \cdots &  *    &  1
\end{matrix} \right),
\]
where $\xi_k$'s are the polynomials of $(a_j,b_j)$ in the Lax matrix.
Since the Toda lattice is isospectral, those polynomials stays constants
even when all of the elements $(a_j, b_j)$ blows up. Then the limit matrix
$L_J$ is obtained
by the limit of the conjugation of $L$ with $x_J\in N^+_J$,
\[
L_J=\lim_{t\to t_J} Ad(x^{-1}_J(t))L(t), \quad {\rm with}\quad
x_J=\left(\begin{matrix}
1       & 0      &\cdots  & \cdots & \cdots &  0 \\
0       & 1      &\cdots  & \cdots & \cdots &  0  \\
\vdots  & \ddots &\ddots  & \cdots & \ddots & \vdots \\
\vdots  & \vdots &\vdots  & x'_J   & \vdots & \vdots \\
\vdots  & \ddots &\ddots  & \cdots & \ddots & \vdots \\
0       & \cdots &\cdots  & \cdots & \cdots &  1
\end{matrix} \right).
\]
Let us now give an example to illustrate the construction:
\begin{Example}
\label{A3example} The $A_3$ Toda lattices:
The Lax matrix is given by
\[
L=\left( \begin{matrix}
b_1 & 1 & 0 & 0 \\
a_1 & b_2 & 1 & 0 \\
0 & a_2 & b_3 & 1\\
0 & 0 & a_3 & b_4
\end{matrix} \right), \quad \sum_{k=1}^4 b_k=0.
\]

The limit matrices $L_J$ are determined as follows:
\begin{itemize}
\item[a)] $J=\{1\}$: Then $\tau_1(t)\sim t_*=(t-t_{\{1\}})$ implies that
$a_1\sim t_*^{-2}, a_2\sim t_*, b_1\sim t_*^{-1}, b_2\sim t_*^{-1}$ and
others are
regular. The limit matrix is then obtained by the limit
$x_{\{1\}}^{-1}Lx_{\{1\}}\to L_{\{1\}}$ as $t_*\to 0$,
\[
L_{\{1\}}=\left(\begin{matrix}
0 & 1 & 0 & 0\\
-\xi_2& \xi_1& 1 & 0\\
\eta_1& 0 &b_3 & 1\\
0 & 0 & a_3 & b_4
\end{matrix} \right), \quad {\rm with} \quad
x_{\{1\}}=\left( \begin{matrix}
1 & 0 & 0 & 0\\
-b_1& 1 & 0  & 0\\
0 & 0 & 1 & 0\\
0 & 0 & 0 & 1
\end{matrix} \right)
\]
where $\xi_1=b_1+b_2, ~\xi_2=b_1b_2-a_1,~ \eta_1=-a_2b_1$ are the parameters
for the divisor
${\mathcal D}_{\{1\}}$.

\item[b)] $J=\{2\}$: With $\tau_2\sim t_*=(t-t_{\{2\}})$, we have $a_1\sim
t_*, a_2\sim t_*^{-2}, a_3\sim t_*, b_2\sim t_*^{-1}, b_3\sim t_*^{-1}$, and
the limit matrix is given by
\[
L_{\{2\}}=\left(\begin{matrix}
b_1 & 1 & 0 & 0\\
0 & 0 & 1 & 0\\
\eta_1& -\xi_2 & \xi_1 & 1\\
0 & \eta_2 & 0 & b_4
\end{matrix} \right), \quad {\rm with} \quad
x_{\{2\}}=\left( \begin{matrix}
1 & 0 & 0 & 0\\
0 & 1 & 0  & 0\\
0 & -b_2 & 1 & 0\\
0 & 0 & 0 & 1
\end{matrix} \right)
\]
where $\xi_1=b_2+b_3,~ \xi_2=b_2b_3-a_2, ~\eta_1=-a_1b_3,~ \eta_2=-a_3b_2$
are the parameters for the divisor ${\mathcal D}_{\{2\}}$.

\item[c)] $J=\{3\}$: This case is similar to the case $J=\{1\}$, and we have
\[
L_{\{3\}}=\left(\begin{matrix}
b_1 & 1 & 0 & 0\\
a_1& b_2 & 1 & 0\\
0 & 0 &0  & 1\\
0 & \eta_1 & -\xi_2 & \xi_1
\end{matrix} \right), \quad {\rm with} \quad
x_{\{3\}}=\left( \begin{matrix}
1 & 0 & 0 & 0\\
0 & 1 & 0  & 0\\
0 & 0 & 1 & 0\\
0 & 0 & -b_3 & 1
\end{matrix} \right)
\]
where $\xi_1=b_3+b_4,~ \xi_2=b_3b_4-a_3,~ \eta_1=-a_2b_4$ are the parameters
for the divisor
${\mathcal D}_{\{3\}}$.

\item[d)] $J=\{1,2\}$: We construct $L_{\{1,2\}}$ from $L_{\{1\}}$ with
$x_{\{1,2\}}$,
\[
L_{\{1,2\}}=\left(\begin{matrix}
0 & 1 & 0 & 0\\
0 & 0 & 1 & 0\\
\xi'_3& -\xi'_2 &\xi_1' & 1\\
\eta_1' & 0 & 0 & b_4
\end{matrix} \right), \quad {\rm with} \quad
x_{\{1,2\}}=\left( \begin{matrix}
1 & 0 & 0 & 0\\
0 & 1 & 0  & 0\\
\xi_2  & -\xi_1 & 1 & 0\\
0 & 0 & 0 & 1
\end{matrix} \right)
\]
where $\xi_1'=\xi_1+b_3, ~\xi_2'=\xi_2+\xi_1b_3, ~\eta_1'=\eta_1+\xi_2b_3$
with $\xi_1, \xi_2, \eta_1$ in $L_{\{1\}}$ are then the parameters for the
divisor
${\mathcal D}_{\{1,2\}}$. This can be of course done with a matrix
$x_{\{2,1\}}$ from
$L_{\{2\}}$.

\item[e)] $J=\{2,3\}$: This is similar to the previous case d), and we have
\[
L_{\{2,3\}}=\left(\begin{matrix}
b_1 & 1 & 0 & 0\\
0 & 0 & 1 & 0\\
0 & 0 & 0 & 1\\
\eta'_1 & \xi_3' & -\xi_2' & \xi_1'
\end{matrix} \right), \quad {\rm with} \quad
x_{\{2,3\}}=\left( \begin{matrix}
1 & 0 & 0 & 0\\
0 & 1 & 0  & 0\\
0 & 0 & 1 & 0\\
0 &\xi_2 & -\xi_1 & 1
\end{matrix} \right)
\]
where $\xi_1'=\xi_1+b_2,~ \xi_2'=\xi_2+\xi_1b_2,~ \eta_1'=\eta_1+\xi_2b_2$
with $\xi_1, \xi_2, \eta_1$ in $L_{\{2\}}$ are then the parameters for the
divisor
${\mathcal D}_{\{2,3\}}$.

\item[f)] $J=\{1,3\}$: We construct the limit matrix $L_{\{1,3\}}$ from
$L_{\{1\}}$
by using $x_{\{1,3\}}=x_{\{3\}}$.
\[
L_{\{1,3\}}=\left(\begin{matrix}
0     & 1    & 0      & 0\\
-\xi_2& \xi_1& 1      & 0\\
0     & 0    & 0      & 1\\
\eta'_1& 0    &-\xi'_2 & \xi'_1
\end{matrix} \right),
\]
where $\xi_1'=b_3+b_4,~ \xi_2'=b_3b_4-a_3,~ \eta'_1=-\eta_1b_3$.

\end{itemize}
\end{Example}

\section{The $A_l$ Toda lattice and the Mumford system} \label{mumfordsect}
In \cite{mumford:84}, Mumford gave a parametrization of the theta divisor
for a hyperelliptic Jacobian with triples of polynomials determined by the
factorization of the corresponding hyperelliptic curve. This is related to the
periodic Toda lattice, but the idea can be also applied to the present case
of nonperiodic Toda lattice on ${\mathfrak g}={\mathfrak sl}(l+1,{\mathbb R})$:

\begin{Definition}
\label{mumfordsystem}
The Mumford system for the spectral curve $F_l(\lambda)={\rm det}(\lambda I-L)$
of degree $l+1$ is the triples of polynomials $(u_d(\lambda), v_d(\lambda),
w_d(\lambda))$ determined by
\[
F_l(\lambda)=u_d(\lambda)w_d(\lambda)+v_d(\lambda), \]
where $u_d$ is a monic polynomial of degree $d$, $v_d$ is a polynomial of
degree less than $d$ with the condition $v_d(\mu_k)=F_l(\mu_k)$ for the roots
of $u_d(\lambda)=0$, and $w_d$ is a monic polynomial of degree $l+1-d$.
\end{Definition}
One can write $u_d$ and $v_d$ in the form,
\[\left\{
\begin{array}{ll}
u_d(\lambda) &=\displaystyle{\prod_{k=1}^d(\lambda-\mu_k)}, \\
v_d(\lambda) &=\displaystyle{\sum_{k=1}^dF_l(\mu_k)\prod_{j\ne k}{\lambda-\mu_
j \over \mu_k-\mu_j}}.
\end{array}
\right.
\]
When $d=l$ (the rank of the matrix), the $\mu$-variables can globally
parametrize the isospectral manifold $Z(\gamma)_{\mathbb R}$ by taking an
explicit
relation with the original variables $(a_k, b_k)$ in $L$, for example,
choose the $l$-th principal minor of $L$ to be $u_l(\lambda)$.
One can also define an integrable system for the Mumford system as
\begin{equation}
\label{mumfordequation}
\left\{
\begin{array}{ll}
\displaystyle{{du \over dt}} &= v ,\\
\displaystyle{{dv \over dt}} &= \displaystyle{u\left[{vw\over
u}\right]_+-vw}, \\
\displaystyle{{dw \over dt}} &=-\displaystyle{\left[{vw \over u}\right]_+},
\end{array}
\right.
\end{equation}
where $[f(\lambda)]_+$ indicates the polynomial part of $f(\lambda)$
(see \cite{mumford:84, vanhaecke:98} for the periodic case).
The integrability is a direct consequence of the isospectrality,
i.e. fixing the curve $F_l(\lambda)=uw+v$. It is also interesting to note that
the system
has a Lax form,
\[ {dM\over dt}=[M, B], \quad {\rm with}\quad M=\left(
\begin{matrix}
h & u\\
w & -h
\end{matrix}\right), \quad B={1\over 2h}\left(
\begin{matrix}
0 & v\\
b & 0
\end{matrix}\right),
\]
where $h^2=v$ and $b=[vw/u]_+$. Then the first equation in
(\ref{mumfordequation}) gives the system,
\[
{d\mu_k\over dt}=-\displaystyle{{F_l(\mu_k)\over \prod_{j\ne k}(\mu_k-\mu_j)}},
\quad k=1,\cdots,d. \]
Using the Lagrange interpolation formula,
\begin{equation}
\nonumber
\displaystyle{\sum_{k=1}^{d}{\mu_k^n \over \prod_{j\ne k}(\mu_k-\mu_j)}=}
\left\{
\begin{array}{ll}
0 \quad {\rm if} ~ n<d-1,\\
1  \quad {\rm if} ~ n=d-1. \\
\end{array}
\right.
\end{equation}
we obtain, after integration,
\[
\displaystyle{\sum_{k=1}^{d}\int_{\mu_0}^{\mu_k}{\lambda^n d\lambda \over
F_l(\lambda)}}=\left\{
\begin{matrix}
c_n  & n< d-1,\\
- t+c_{d-1} & n=d-1. \\
\end{matrix}
\right.
\]
with some constants $\mu_0$ and $c_k, k=1,\cdots, d-1$.
In particular, the system with $d=1$ gives
\[ \displaystyle{{d\mu_1\over dt}=-F_l(\mu_1)},\]
whose solution has $l+1$ fixed points at $\mu_1=\lambda_k$ for
$k=1,\cdots,l+1$,
and blows up when $\mu_1> \underset{k}{{\rm max}}(\lambda_k) $ or $\mu_1<
\underset{k}{{\rm min}}(\lambda_k)$.
One can also show the following Proposition on the topology of
certain 1-dimensional Painlev\'e divisors $\Theta_{J}(A_k)$ of the $A_k$
Toda lattice:
\begin{Proposition}
\label{1dpainlevedivisors}
Let $J_{k-1}\subset\{1,\cdots,k\}$ be either $\{1,\cdots,k-1\}$ or
$\{2,\cdots,k\}$.
Then the Painlev\'e divisors $\Theta_{J_{k-1}}(A_k)$ are all homeomorphic
to circle, i.e.
\[ \Theta_{J_{k-1}}(A_k) \cong  S^1,\quad {\rm for} \quad k=1,2,\cdots,\]
where $J_0=\emptyset$.
\end{Proposition}
\begin{Proof}
Since the homeomorphism between the divisors with $J=\{1,\cdots,k-1\}$ and
$J=\{2,\cdots,k\}$ is obvious, we consider the case with
$J=\{1,\cdots,k-1\}$. In this case, the limit matrix
has the form,
\[ L_J=\left(\begin{matrix}
0      & 1    & 0    & \cdots & \cdots & 0 \\
0      & 0    & 1    & \cdots & \cdots & 0 \\
\vdots &\ddots&\ddots&\ddots  &\ddots  &\vdots\\
0      &\ddots&\ddots&\ddots  &  1     & 0  \\
(-1)^{k-1}\xi_k&\cdots&\cdots&\cdots&\xi_1& 1 \\
\eta   & 0    & \cdots&\cdots& 0  & b_{k+1}
\end{matrix} \right),\]
where $\xi_i$ are the coefficients of the polynomial $|\lambda
I-L'|=\lambda^{k}+
\sum_{i=1}^k (-1)^i \xi_i\lambda^{k-i}$ with $L'$ given by the first
$k\times k$
part of the Lax matrix $L$, and $\eta=-a_{k}b_1\cdots b_{k-1}$ (in the limit
$t\to t_J$).
Then from the Mumford system $F_k(\lambda):=|\lambda I-L_J|=u_1w_1+v_1$, we
have
\[ \eta=-F_k(\mu_1), \quad {\rm with}\quad \mu_1=b_{k+1},\]
where $v_1=-\eta$. This indicates that the Painlev\'e
divisor
${\mathcal D}_{J}(A_k)$ has just one connected component of $\mathbb R$, and
adding
the highest divisor $\Theta_{\{1,\cdots,k\}}(A_k)$ we see that the closure
$\Theta_J(A_k)$ is homeomorphic to $S^1$. This completes the proof.
\end{Proof}
We can also show the following on higher dimensional divisors,
\begin{Proposition}
\label{divisorequivalenceD}
Let $J_{n}^k \subset\{1,\cdots,k+n\}$ be either $\{1,\cdots,n\}$ or
$\{k+1,\cdots,k+n\}$.
 The Painlev\'e divisors ${\mathcal D}_{J_n^k}(A_{k+n})$ are all
homeomorphic to
the top cell of the $A_k$ Toda lattice, i.e.
\[{\mathcal D}_{\emptyset}(A_k)\cong {\mathcal D}_{J_{n}^k}(A_{k+n}), \quad
{\rm for}\quad n\ge 1.\]
\end{Proposition}
\begin{Proof}
Let $J$ be $\{1,\cdots,n\}$. Then the limit matrix $L_J$ is given by
\[L_J=\left(\begin{matrix}
A_1 & A_2 \\
A_3 & A_4
\end{matrix}\right),\]
where $A_1$ is the $(n+1)\times(n+1)$ companion matrix
of the corresponding block in the matrix $L$, $A_2$ is the
$(n+1)\times k$ matrix
having zero entries except 1 at the bottom left corner (i.e. $(k,1)$-entry),
$A_3$ is the
$k\times (n+1)$ matrix having zero entries except $\eta$ at the top left
corner ($(1,1)$-entry), and $A_4$ is the $k\times k$ submatrix of Lax
matrix,
\[A_4=\left(
\begin{matrix}
b_{n+2} & 1       & \cdots & \cdots & 0  \\
a_{n+2} & b_{n+3} & 1      &\cdots  & 0 \\
\vdots  & \ddots  & \ddots &\ddots  & \vdots \\
0       & \cdots  & \cdots & b_{k+n}& 1  \\
0       &\cdots   &  \cdots& a_{k+n}&b_{k+n+1}
\end{matrix}\right).\]
Then from the factorization of $F_{k+n}(\lambda)=|\lambda
I-L_J|=u_k(\lambda)w_k(\lambda)+v_k(\lambda)$,
we have the Mumford system,
\[
u_k=|\lambda I-A_4|, \quad v_k=-\eta |\lambda I-B_4| , \quad w_k=|\lambda
I-A_1|,\]
where $B_4$ is the $(k-1)\times (k-1)$ submatrix of $A_4$ by deleting the first
row and column vectors. Thus we have
\[
\eta |\mu_j I-B_4|=-F_{k+n}(\mu_j), \quad {\rm for}\quad j=1,\cdots,k,\]
The left-hand side of this equation has the same form for all the cases with
fixed $k$,
and the right-hand side
gives a real one-dimensional affine curve for each $\mu_j\in {\mathbb R}$ of
degree $k+n$.
This implies that all the divisors ${\mathcal D}_{J_n^k}(A_{k+n})$ have the
same parametrization,
so that they are all homeomorphic.
\end{Proof}
Since the boundaries of
each ${\mathcal D}_{J_n^k}(A_{k+n})$ seems to have the same structure for
$n\ge 1$,
we expect
\begin{conjecture}
\label{divisorequivalence}
The Painlev\'e divisors $\Theta_{J_n^k}(A_{k+n})$ for $n\ge 1$ are all
homeomorphic, i.e.
\[\Theta_{J_1^k}(A_{k+1})\cong \cdots \cong \Theta_{J_n^k}(A_{k+n}).\]
\end{conjecture}

\section{Examples for rank 2 and 3}
\label{rank2example}
\subsection{The $A_2$-Toda lattice}\label{A2todalattice}
The Lax matrix is a $3\times 3$ matrix given by
\[
L=\left(
\begin{matrix}
b_1 &1 &0\\
a_1 &b_2&1\\
0 & a_2& b_3
\end{matrix}
\right), \quad {\rm with}\quad \sum_{k=1}^3b_k=0,
\]
 and the spectral curve
$F_2(\lambda)={\rm det}(\lambda I-L)$ is
\[ F_2(\lambda)=\lambda^3+I_1\lambda-I_2,\]
where the Chevalley invariants $I_k(L)$ are given by
\[
I_1(L)= b_1b_2+b_2b_3+b_1b_3-a_1-a_2, \quad I_2(L)=b_1b_2b_3-a_1b_3-a_2b_1.
\]
To parametrize the isospectral manifold $Z(\gamma)_{\mathbb R}$, we consider
for example the following Mumford system with the choice of the triples,
\[
\left\{
\begin{array}{ll}
u_2(\lambda) &= \prod_{k=1}^2(\lambda-\mu_k) =\left|
\begin{matrix}
\lambda-b_2 &-1\\
-a_2 & \lambda-b_3
\end{matrix} \right| ,\\
v_2(\lambda) &= \displaystyle{F_2(\mu_1){\lambda-\mu_2 \over
\mu_1-\mu_2}+F_2(\mu_2)
{\lambda-\mu_1 \over \mu_2-\mu_1}}, \\
w_2(\lambda) &= \displaystyle{{1\over
u_2(\lambda)}(F_2(\lambda)-v_2(\lambda))=\lambda+w_0}.
\end{array}
\right.
\]
Then in terms of $(\mu_1,\mu_2)$ the Chevalley invariants are given by
\[ I_1=-(\mu_1+\mu_2)^2+\mu_1\mu_2-a_1,\quad
I_2=-\mu_1\mu_2(\mu_1+\mu_2)-a_1b_3,\]
 which leads to
\[ a_1(\mu_k-b_3)=-F_2(\mu_k),\quad k=1,2.\]

\begin{figure}[tb]
\includegraphics{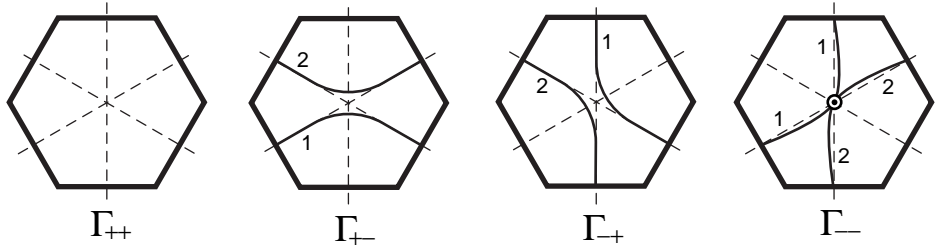}
\caption{The $A_2$ hexagons $\Gamma_{\epsilon_1\epsilon_2}$. The numbers 1 and
2 mark the Painlev\'e
divisors $\Theta_{\{1\}}$ and $\Theta_{\{2\}}$. The center point (double
circle) in
$\Gamma_{--}$
indicates the divisor $\Theta_{\{1,2\}}$.}
\label{A2:fig}
\end{figure}

 Also from (\ref{mumfordequation}), we have the Toda flow in the variable
$(\mu_1,\mu_2)$,
\[
\displaystyle{{d\mu_k \over dt}}=(-1)^k\displaystyle{{F_2(\mu_k)\over
\mu_1-\mu_2}}, \quad k=1,2,
\]
which is also obtained by setting
$a_1(\mu_k-b_3)=(-1)^{k+1}(\mu_1-\mu_2)d\mu_k/dt$.
The system has 6 fixed points with $(\mu_1,\mu_2)=(\lambda_i,\lambda_j), ~
1\le i\ne j\le 3$, and for each set of the signs $(\epsilon_1,\epsilon_2)$ with
$\epsilon_i={\rm sign}(a_i(0))$ the integral manifold gives a hexagon,
denoted by
$\Gamma_{\epsilon_1 \epsilon_2}$ as in Fig.\ref{A2:fig}. In particular,
one can easily see that there is no blow-up in $\Gamma_{++}$ (note that
$I_1(L)=\gamma_1$ makes all the variables be bounded,
 if both $a_1$ and $a_2$ are positive). Those four hexagons are glued
together along
 with their boundaries according to the standard action of the Weyl group
$S_3$ on the signs $(\epsilon_1,\epsilon_2)$, and the compactified manifold
is topologically equivalent to
a connected sum of two Klein bottles ${\mathbb K}$ \cite{kodama:98}. This can
be seen by
counting the Euler characteristic, $6(vertices)-12(edges)+4(hexagons)=-2$ and the
nonorientability (see \cite{casian:99} for the general argument on the
compactification based on the Weyl group action).

 The Painlev\'e divisor ${\mathcal D}_{\{1\}}$ corresponding to $\tau_1=0$
can be
parametrized by
 the limit matrix,
\begin{equation}
\label{theta1}
L_{\{1\}}=\left(
\begin{matrix}
0 & 1 &  0\\
-\xi_2 &\xi_1 &1 \\
\eta_1 & 0 & b_3
\end{matrix}\right),
\end{equation}
where $\xi_1=b_1+b_2, \xi_2=b_1b_2-a_1$ and $ \eta_1=-a_2b_1$. The matrix
$L_{\{1\}}$
is obtained by the limit,
\[
L_{\{1\}}=\lim_{t\to t_{1}} x_{\{1\}}^{-1}(t)L(t)x_{\{1\}}(t), \quad {\rm
with} \quad
x_{\{1\}}(t)=\left(
\begin{matrix}
1 & 0 & 0\\
-b & 1 & 0 \\
0  & 0 & 1
\end{matrix}
\right).
\]
Then the spectral curve $F_2(\lambda)$ gives the algebraic relations (the
Chevalley invariants),
\[ \xi_1+b_3=0,~ I_1(L_{\{1\}})=\xi_2+\xi_1b_3, ~
I_2(L_{\{1\}})=\eta_1+\xi_2b_3,\]
which leads to
\[
\begin{array}{ll}
{\mathcal D}_{\{1\}} &= \left\{ (\xi_1,\xi_2,\eta_1,b_3)\in {\mathbb R}^4~:~
\xi_1=-b_3,
I_1=\gamma_1,
I_2=\gamma_2~\right\} \\
{} &= \left\{(\eta_1, b_3) \in {\mathbb R}^2~:~
\eta_1=-F_2(b_3)=-b_3^3-\gamma_1b_3+\gamma_2\right\}.
\end{array}
\]
We thus show that the closure of ${\mathcal D}_{\{1\}}$ is homeomorphic to a
circle $S^1$, and it intersects with three subsystems corresponding to
$(a_2=0,b_3=\lambda_k)$ for $k=1,2,3$. The Mumford equation
(\ref{mumfordequation}) can be used to provide a dynamics on
${\mathcal D}_{\{1\}}$ with $\mu_1=b_3$ and $\eta_1=d\mu_1/dt$,
\[\displaystyle{{d\mu_1\over dt}=-F_2(\mu_1).}\]
In Figure \ref{A2:fig}, $\Theta_{\{1\}}$ is shown as a curve with the
label \lq\lq 1\rq\rq.
The $\Theta_{\{2\}}$ has the similar structure.
Thus we obtain:
\begin{Proposition}
\label{A2topology}
The compactified manifold ${\tilde Z}(\gamma)_{\mathbb R}$ and the
Painlev\'e divisor have the
following topology,
\[ {\tilde Z}(\gamma)_{\mathbb R}=\Theta_{\emptyset}\cong {\mathbb
K}~\sharp~{\mathbb K}~,
\quad \Theta_{\{1\}}\cong\Theta_{\{2\}}\cong S^1.\]
\end{Proposition}
We also
note by taking out the divisors $\Theta_{\{1\}}$ and $\Theta_{\{2\}}$ from
${\tilde Z}(\gamma)_{\mathbb R}$ that
the top cell ${\mathcal D}_{\emptyset}={\tilde Z}(\gamma)_{\mathbb R}\cap
N^-B^+/B^+$ is diffeomorphic to
a torus ${\mathbb T}$ with a hole of a disk ${\mathbb D}$, i.e.
\[ {\mathcal D}_{\emptyset}\cong {\mathbb T}\setminus {\mathbb D}. \]

\subsection{The $C_2$ Toda lattice}
Since the $B_2$ Toda lattice has the same structure as the $C_2$ case, we
discuss
only the latter one.
The Lax matrix for $C_2$ Toda lattice is given by a $4\times 4$ matrix,
\[
L=\left( \begin{matrix}
b_1 & 1 & 0 & 0 \\
a_1 &b_2&1 & 0\\
0 & 2a_2 & -b_2 & 1\\
0 &0 & a_1 & - b_1
\end{matrix} \right)
\]
whose spectral curve $F_2(\lambda)={\rm det}(\lambda I-L)$ is
\[ F_2(\lambda)=\lambda^4-I_1\lambda^2+I_2\]
with the Chevalley invariants $I_k(L)$,
\[I_1=b_1^2+b_2^2+2a_1+2a_2, ~ I_2=(b_1b_2-a_1)^2+2b_1^2a_2.\]
The corresponding polytope $\Gamma_{\epsilon_1 \epsilon_2}$
with the signs $\epsilon_k={\rm sign}(a_k)$ is given by a octagon with eight
vertices
associated with the fixed point of the system, $a_1=a_2=0$. Those vertices
are expressed as
$(b_1,b_2)=(\sigma_1\lambda_i,\sigma_2\lambda_j)$ for $\sigma_k\in\{\pm\},
i\ne j\in\{ 1,2\}$. Gluing those octagons along their boundaries, we find
that the compactified manifold
${\tilde Z}(\gamma)_{\mathbb R}$ is topologically equivalent to a connected
sum of
three Klein bottles ${\mathbb K}$. Again just count the Euler characteristic,
$8(vertices)-16(edges)+4(octagons)=-4$, and the nonorientability leads to
the result.

\begin{figure}[tb]
\includegraphics{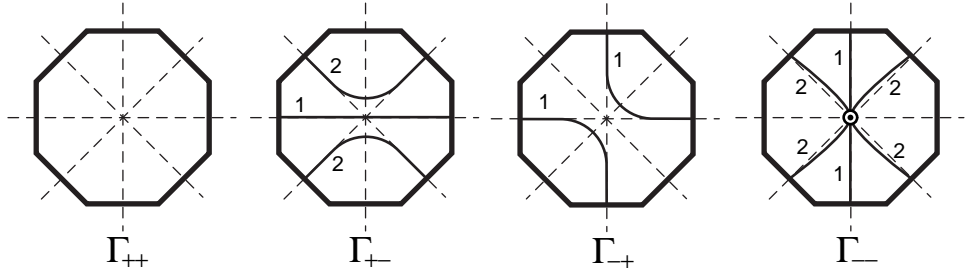}
\caption{The $C_2$ octagons $\Gamma_{\epsilon_1\epsilon_2}$ and the Painlev\'e
divisors
$\Theta_{\{1\}}, \Theta_{\{2\}}$ and $\Theta_{\{1,2\}}$.}
\label{B2:fig}
\end{figure}

The Painlev\'e divisor $\Theta_{\{1\}}$ is now parametrized by the limit matrix
\[
L_{\{1\}}=\left(
\begin{matrix}
0 & 1 & 0 & 0\\
-\xi_2 & \xi_1 & 1 & 0\\
0 & 0 & 0 & 1 \\
\eta_1 & 0 & -\xi_2 & -\xi_1
\end{matrix}
\right)
\]
where $\xi_1=b_1+b_2, \xi_2=b_1b_2-a_1$ and $\eta_1=-2a_2b_1^2$.
Then the Chevalley invariants $I_k(L)$ are expressed by
\[I_1=\xi_1^2-2\xi_2,\quad I_2=\xi_2^2-\eta_1 \]
 from which we obtain
\[ \eta_1={1\over 4}((\xi_1^2-I_1)^2-4I_2).\]
This implies that the $\Theta_{\{1\}}$ is homeomorphic to $S^1$ and intersects
with four subsystems corresponding to $\xi_1=\sigma(\lambda_1\pm\lambda_2)$
with $\sigma\in\{\pm 1\}$ and with the divisor $\Theta_{\{2\}}$ in
$\Gamma_{--}$ (see
Figure \ref{B2:fig}),

Unlike the case of $A_2$ Toda lattice, the divisor $\Theta_{\{2\}}$ has a
different
structure. The corresponding limit matrix $L_{\{2\}}$ is given by
\[
L_{\{2\}}=\left(
\begin{matrix}
b_1 & 1 & 0 & 0\\
0  &  0 & 1 & 0\\
\eta_2 & \xi_3 & 0 & 1\\
0   &  - \eta_2 & 0 & -b_1
\end{matrix} \right)
\]
where $\xi_3=b_2^2+2a_2, \eta_2=a_1b_2$.
The invariants $I_k$ are then given by
\[I_1=b_1^2+\xi_3, \quad I_2= \xi_3b_1^2-2\eta_2b_1, \]
and we obtain
\[\eta_2=-{1\over b_1}F_2(b_1). \]
Because of the singularity in this equation at $b_1=0$, the $\Theta_{\{2\}}$ is
shown to be
homeomorphic to a figure eight, where each circle intersects two subsystems
corresponding to either $b_1=|\lambda_k|$ or $b_1=-|\lambda_k|$ with $k=1,2$.
The node of the figure eight corresponds to the divisor $\Theta_{\{1,2\}}$
(see Figure \ref{B2:fig}). We thus obtain,
\begin{Proposition}
\label{C2topology}
The topology of the isospectral manifold of $C_2$ and the Painlev\'e divisor
is given by
\[{\tilde Z}(\gamma)_{\mathbb R}\cong {\mathbb K}~\sharp~{\mathbb
K}~\sharp~{\mathbb K}~, \quad
\Theta_{\{1\}}\cong S^1,\quad \Theta_{\{2\}}\cong S^1\vee S^1.\]
\end{Proposition}
The $B_2$ Toda lattice has the same structure, but $\Theta_{\{1\}}$ and
$\Theta_{\{2\}}$
have the opposite structure.

\subsection{The $G_2$ Toda lattice}
We use the following one for the Lax matrix,
 \[
L= \left(
\begin{matrix}
b_1    &    1     &   0    & \cdot   & \cdot   & \cdot   &  0 \\
a_1    & b_2  &  1     &  0      & \cdot   & \cdot   &  0 \\
0      &  a_2     & b_1-b_2 & 1     &  0      & \cdot   &  0 \\
0      &    0     & 2a_1   & 0       & 1       & 0       &  0  \\
0      & \cdot    &  0     & 2a_1  & -b_1+b_2 & 1      &  0 \\
0      & \cdot    & \cdot  &  0      &  a_2    & -b_2   &  1 \\
0      & \cdot    & \cdot  &  \cdot  &  0      & a_1     & -b_1 \\
\end{matrix}
\right).
\]
The spectral curve is then given by
\[ F_2(\lambda)=\lambda (\lambda^2(\lambda^2 +I_1)^2+I_2), \]
where $I_1$ and $I_2$ are the Chevalley invariants given by homogeneous
polynomials of $(a_1,\cdots, b_2)$. Each polygon
$\Gamma_{\epsilon_1,\epsilon_2}$ in the isospectral manifold has 12 vertices
corresponding to $a_1=a_2=0$ which is also the order of the
Weyl group. Those polygons are glued to
obtain the compactified manifold which is topologically equivalent to a sum
of five Klein bottles. The Euler characteristic is
$12(vertices)-24(edges)+4(polygons)=-8$.

\begin{figure}[tb]
\includegraphics{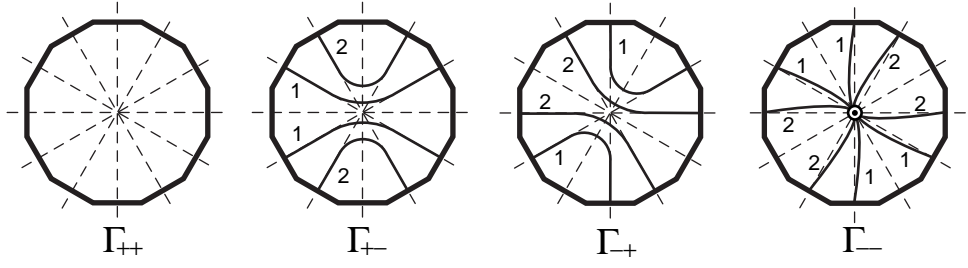}
\caption{The $G_2$ polygons $\Gamma_{\epsilon_1\epsilon_2}$ with the Painlev\'e
divisors
$\Theta_{\{1\}}, \Theta_{\{2\}}$ and $\Theta_{\{1,2\}}$.}
\label{G2:fig}
\end{figure}

The Painlev\'e divisor ${\mathcal D}_{\{1\}}$ is parametrized by the limit
matrix $L_{\{1\}}$,
\[
L_{\{1\}}=\left(\begin{matrix}
0      &    1     &   0    & \cdot   & \cdot   & \cdot   &  0 \\
-\xi_2 & \xi_1    &  1     &  0      & \cdot   & \cdot   &  0 \\
0      &  0       & 0      & 1       &  0      & \cdot   &  0 \\
0      &    0     & 0      & 0       & 1       & 0       &  0  \\
\eta   &    0     &  0     &\xi_1^2-4\xi_2 & 0 & 1       &  0 \\
0      &    0     & 0      &  0      &  0      & 0       &  1 \\
0      &    0     & -\eta  &  0      &  0      & -\xi_2  & -\xi_1 \\
\end{matrix}
\right),
\]
where $\xi_1=b_1+b_2,~\xi_2=b_1b_2-a_1$ and $\eta=2b_1a_1a_2$
in the limit $t\to t_{\{1\}}$ with $\tau_1(t)\sim t-t_{\{1\}}$.
Here we have used the conjugating matrix $x_{\{1\}}$ as,
\[
x_{\{1\}}=\left(\begin{matrix}
1      &    0     &   0    & \cdot   & \cdot   & \cdot   &  0 \\
-b_1   &   1      &   0    &  0      & \cdot   & \cdot   &  0 \\
0      &  0       &   1    &  0      &  0      & \cdot   &  0 \\
0      &    0     & b_2-b_1&  1      &  0      & 0       &  0  \\
0      &    0     &  -2a_1 & b_2-b_1 &  1      &  0      &  0 \\
0      &    0     & 0      &  0      &  0      &  1      &  0 \\
0      &    0     & 0      &  0      &  0      & -b_1    &  1 \\
\end{matrix}
\right) ,
\]
which can be obtained from the structure of $\tau$-functions in
(\ref{tau}).
Then the invariants $I_1, I_2$ are given by
\[
I_1=3\xi_2-\xi_1^2, \quad I_2= (4\xi_2-\xi_1^2)\xi_2^2+2\eta\xi_1.
\]
Eliminating $\xi_2$, we obtain
\[
\eta=\displaystyle{{1\over 2\xi_1}\left(-{1\over
27}(\xi_1^2+I_1)^2(\xi_1^2+4I_1)+I_2\right)}.\]which has two connected
components, and each component intersects three times with the
boundaries of the polytopes $\Gamma_{+-},\Gamma_{-+}$ and $\Gamma_{--}$ (see
Figure \ref{G2:fig}).

The limit matrix corresponding to the Painlev\'e divisor ${\mathcal
D}_{\{2\}}$ is given by
\[L_{\{2\}}=\left(\begin{matrix}
\xi_1  &    1     &   0    & \cdot   & \cdot   & \cdot   &  0 \\
 0     & 0        &  1     &  0      & \cdot   & \cdot   &  0 \\
\eta   & -\xi_2   & \xi_1  & 1       &  0      & \cdot   &  0 \\
0      &  -2\eta  & 0      & 0       & 1       & 0       &  0  \\
0      &    0     &  0     &  0      & 0       & 1       &  0 \\
0      &    0     & 0      & 2\eta   & -\xi_2  & -\xi_1  &  1 \\
0      &    0     &  0     &  0      & -\eta   &  0      & -\xi_1 \\
\end{matrix}
\right)
\]
with the conjugating matrix $x_{\{2\}}$,
\[
x_{\{2\}}=\left(\begin{matrix}
1      &    0     &   0    & \cdot   & \cdot   & \cdot   &  0 \\
0      &   1      &   0    &  0      & \cdot   & \cdot   &  0 \\
0      &  -b_2    &   1    &  0      &  0      & \cdot   &  0 \\
0      &    0     &  0     &  1      &  0      & 0       &  0  \\
0      &    0     &  0     &  0      &  1      &  0      &  0 \\
0      &    0     & 0      &  0      & b_1-b_2 &  1      &  0 \\
0      &    0     &  0     &  0      &  0      &  0      &  1 \\
\end{matrix}
\right).
\]
Here the new variables are $\xi_1=b_1, ~\xi_2=b_2(b_1-b_2)-a_2$ and
$\eta=a_1b_2$ in the corresponding limit with $\tau_2(t)\to 0$.
With those variables, the invariants are
\[
I_1=\xi_2-\xi_1^2, \quad
I_2=3\eta^2-2\xi_1\eta(\xi_2+2\xi_1^2)-\xi_1^2\xi_2^2, \]
 from which we have two curves $\eta=\eta_{\pm}(\xi_1)$,
\[
\eta_{\pm}=\displaystyle{{1\over 3}\left(\xi_1(3\xi_1^2+I_1)\pm \sqrt{2\xi_1^2
(5\xi_1^4+4I_1\xi_1^2+I_1^2)+I_2}\right)}. \]
Those curves indicate that there are two connected components of the divisor
${\mathcal D}_{\{2\}}$ and each component has three intersections with the
subsystems. Topologically then the divisors ${\mathcal D}_{\{1\}}$ and
${\mathcal D}_{\{2\}}$ are the same, and adding the divisor ${\mathcal
D}_{\{1,2\}}$
one can conclude that the closure of both divisors are topologically
equivalent to
a figure eight. Thus we have
\begin{Proposition}
\label{G2topology}
The topology of the $G_2$ Toda isospectral manifold and the divisor is given by
\[{\tilde Z}(\gamma)_{\mathbb R}\cong \overbrace{{\mathbb K}~\sharp\cdots
\sharp~{\mathbb K}}^{5}~, \quad \Theta_{\{1\}}\cong \Theta_{\{2\}}\cong S^1\vee
S^1.\]
\end{Proposition}

\subsection{The $A_3$ Toda lattice}
In the example \ref{A3example}, we gave the limit matrices for the
Painlev\'e divisors.
Here we discuss the topology of the divisors by computing explicitly the
isospectral
sets of those matrices, i.e.
\[
F_3(\lambda)=\lambda^4+I_1\lambda^2 - I_2\lambda +I_3,\]
where the Chevalley invariants $I_k(L_J), k=1,2,3$ are now expressed in
terms of the parameters in the limit matrices. Here we use the same
parametrizations in Example
\ref{A3example}:
\begin{itemize}
\item[a)] $J=\{1\}$: We take the polynomial $u_2(\lambda)$ in the Mumford
system as
$u_2(\lambda)=\left|\begin{matrix}
\lambda-b_3 & -1\\
-a_3 & \lambda-b_4 \end{matrix}\right|$, i.e. $\mu_1+\mu_2=b_3+b_4=-\xi_1,
~\mu_1\mu_2=b_3b_4-a_3$.
Then the Chevalley invariants are given by
\[\left\{\begin{array}{ll}
I_1&=\xi_2-(\mu_1+\mu_2)^2+\mu_1\mu_2,\\
 I_2&=\eta_1-\mu_1\mu_2(\mu_1+\mu_2)+\xi_2(\mu_1+\mu_2), \\
I_3&=\xi_2\mu_1\mu_2+\eta_1b_4. \end{array}\right.\]
Eliminating $\xi_2$, we find
\[\eta_1(\mu_k-b_4)=-F_3(\mu_k), \quad k=1,2. \]
As was shown in Proposition \ref{divisorequivalenceD}, comparing this with
the top cell
of the $A_2$ Toda lattice in Subsection \ref{A2todalattice}, one can
see
\[{\mathcal D}_{\{1\}}\cong {\mathbb T}\setminus {\mathbb D}.\]
\item[b)] $J=\{2\}$: We take $u_2(\lambda)=(\lambda-b_1)(\lambda-b_4)$, i.e.
$\mu_1=b_1, \mu_2=b_4$. Then we have, using $\xi_1=-(\mu_1+\mu_2)$,
\[\left\{\begin{array}{ll}
I_1 &= \mu_1\mu_2-(\mu_1+\mu_2)^2+\xi_2,\\
I_2 &= \xi_2(\mu_1+\mu_2)-\mu_1\mu_2(\mu_1+\mu_2)+\eta_1+\eta_2, \\
I_3 &= \mu_1\mu_2\xi_2+\mu_1\eta_2+\mu_2\eta_1,
\end{array}\right.\]
which lead to
\[ \displaystyle{\eta_k=(-1)^k{F_3(\mu_k)\over \mu_1-\mu_2}}.\]
\item[c)] $J=\{3\}$: This case is similar to the one with $J=\{1\}$, and we
have the same
formulae of the Chevalley invariants in the variables $\mu_1,\mu_2$ which
are defined as
$u_2(\lambda)=\left|\begin{matrix}
\lambda-b_1 & -1\\
-a_3 & \lambda-b_2 \end{matrix}\right|$, i.e. $\mu_1+\mu_2=b_1+b_2=-\xi_1,
~\mu_1\mu_2=b_1b_2-a_1$.
\item[d)] $J=\{1,2\}$: Here we take $u_1(\lambda)=\lambda-b_4$ for the
Mumford system, i.e.
$\mu_1=b_4$ and $v_1=-\eta'_1$. Then the Chevalley invariants are
\[ I_1=\xi'_2-\mu_1^2,\quad I_2=\xi'_3+\xi'_2\mu_1, \quad
I_3=\xi'_3\mu_1-\eta'_1,\]
and from $v_1=-\eta'_1$, we obtain
\[\eta'_1=-F_3(\mu_1),\]
which implies that ${\mathcal D}_{\{1,2\}}$ intersects with four boundaries
of the polytopes,
and the closure, $\Theta_{\{1,2\}}$ is homeomorphic to a circle.
\item[e)] $J=\{2,3\}$: We get exactly the same result as the previous case
with $\mu_1=b_1$.
\item[f)] $J=\{1,3\}$: The Chevalley invariants are given by
\[I_1=\xi_1\xi'_1+\xi_2+\xi'_2, \quad I_2=\xi_1\xi'_2+\xi_2\xi'_1, \quad
I_3=\xi_2\xi'_2-\eta'_1.\]
Using $\xi_1+\xi'_1=0$ and eliminating $\xi'_1,\xi'_2$, we obtain
\[\displaystyle{\eta'_1={1\over 4\xi_1^2}\left(
\xi_1^2(\xi_1^2+I_1)^2-I_2^2\right)-I_3}.\]
This equation indicates that ${\mathcal D}_{\{1,3\}}$ has two connected
components, each of which intersects with three boundaries of the polytopes.
Each boudary corresponds to a point $(\eta'_1=0, \lambda_i+\lambda_j)$ for
$i\ne j$.
Then we can see
\[\Theta_{\{1,3\}}\cong S^1\vee S^1.\]
\end{itemize}
\begin{figure}[tb]
\includegraphics{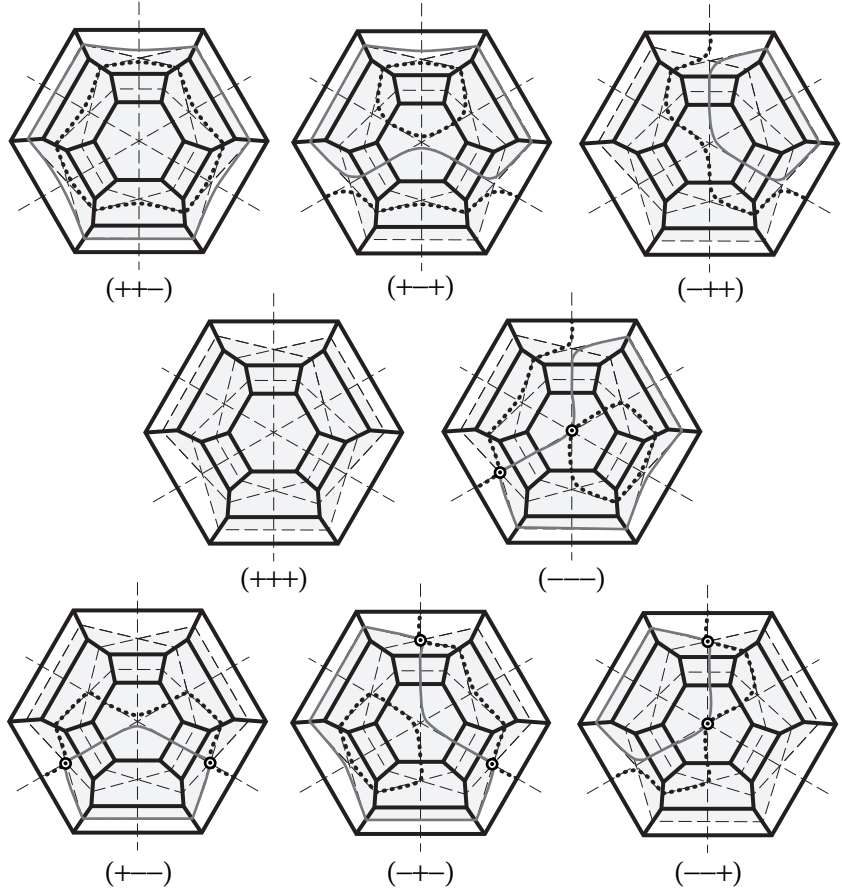}
\caption{The $A_3$ polytopes $\Gamma_\epsilon$ marked by
$\epsilon=(\epsilon_1\epsilon_2\epsilon_3)$ and the Painlev\'e divisors
$\Theta_{\{1 \} }$ (the solid grey curves), $\Theta_{\{2 \} }$ (the dotted
curves) and $\Theta_{\{1,2\}}$ (the double circles).}
\label{A3:fig}
\end{figure}
The results are summarized in Figure \ref{A3:fig} where the Painlev\'e divisors
$\Theta_{\{1\}}, \Theta_{\{2\}}$ and $\Theta_{\{1,2\}}$ are shown as the
solid grey
curves, the dotted curves and the double circles. The $\Theta_{\{3\}}$ has a
similar structure to the $\Theta_{\{1\}}$. One can see from this Figure that
each portion of the $\Theta_{\{1\}}$ on a $\Gamma_{\epsilon}$ is
homeomorphic to either hexagon or octagon,
and we have 4 hexagons in $\Gamma_{\epsilon}$ with
$\epsilon=(++-),(-++),(+--),(--+)$, and
3 octagons with $\epsilon=(+-+),(-+-),(---)$. Then the Euler characteristic
can be computed as follows: The total number of vertices are given by
$12=(4\times 6+3\times 8)/4$ by
identifying 4 vertices of the polygons, the edges are $24=(4\times 6+3\times
8)/2$
in total, and we have 7 faces, i.e. the Euler characteristic is
$12-24+7=-5$. One can also
see the non-orientability of the divisor, so that the $\Theta_{\{1\}}$ is
topologically
equivalent to a connected sum of 7 real projective planes ${\mathbb P}$ (or
3 Klein bottles plus a projective plane). For $\Theta_{\{2\}}$, we have 4
squares and 4 hexagons. However
two squares in $\Gamma_{---}$ are attached at a point of the divisor
$\Theta_{\{1,2,3\}}$,
and thus the $\Theta_{\{2\}}$ gives a singular variety. By detaching those
two squares,
one can compute the Euler characteristic in the same way as above, and we
obtain
$12-24-10=-2$. This shows that the desingularized variety of
$\Theta_{\{2\}}$ is homeomorphic to the compactified manifold ${\tilde
Z}(\gamma)_{\mathbb R}$ for the $A_2$ Toda lattice
(in the next section we give a further discussion on the desingularization
in Lie theoretic point of view).
Thus we have
\begin{Proposition}
\label{A3divisors}
The Painlev\'e divisors for the $A_3$ Toda lattice have the following topology,
\[
\begin{array}{ll}
&\Theta_{\{1\}} \cong  \Theta_{\{3\}}  \cong  {\mathbb K}~\sharp~{\mathbb
K}~\sharp~{\mathbb K}~
\sharp~{\mathbb P}, \quad  \quad \Theta_{\{2\}}^{\vee} \cong  {\mathbb
K}~\sharp~{\mathbb K},\\
&\Theta_{\{1,2\}} \cong \Theta_{\{2,3\}}\cong S^1,\quad \quad
\Theta_{\{1,3\}}\cong S^1\vee S^1 ,
\end{array}
\]
where $\Theta_{\{2\}}^{\vee}$ is the desingularization of $\Theta_{\{2\}}$
by a resolution at the divisor $\Theta_{\{1,2,3\}}$.
\end{Proposition}
The singular structure on the divisor $\Theta_{\{1,3\}}$ has been also found
in the case of periodic Toda lattice \cite{kodama:02}.

\section{An algebraic version of the Painlev\'e divisor}
\label{negativesection} 

Here we discuss the Painlev\'e divisor in the
framework of the Lie theory. We first review and summarize some Lie
theoretic notation.
\subsection{Notations and Definitions}
\begin{Notation}
\label{standardnot}
Lie algebras:
Recall that  $\mathfrak g$ denote a real
split semisimple Lie algebra of rank $l$
and we are fixing a split Cartan subalgebra ${\mathfrak h}$ with root system
$\Delta $, a positive
system $\Delta_{+}$ determining the Borel subgroup $B^{+}$ of $G$. The
corresponding set of simple roots is $\Pi:=\{ \alpha_i : i=1, \cdots ,l \}$
as in Section \ref{chev} where we just denoted $\Pi=\{k=1,\cdots,l\}$.
\end{Notation}

The Weyl group $W$ is thus generated by the simple reflections
$s_{\alpha_i}$, $i=1, \cdots ,l$. For any $S\subset \Pi$, we define the
subgroup generated by
$S$,
\[
W_S=\langle ~s_{\alpha_i}~:~\alpha_i\in S~\rangle
\]
This is the Weyl group of a parabolic Lie subgroup and it is standard to
refer to $W_S$ as a {\it parabolic subgroup} of $W$.

\begin{Notation}
Lie groups:
\label{standardG}
We let $G_{\mathbb C}$ denote the connected adjoint Lie group with Lie
algebra $\mathfrak g_{\mathbb C}$
and $G$ the connected Lie subgroup correspondintg to $\mathfrak g$.
 Denote by ${\tilde G}$ the
Lie group $\left\{g\in G_{\mathbb C}~:~ Ad(g)\mathfrak g \subset \mathfrak g
\right\}.$
A split Cartan of ${\tilde G}$
with Lie algebra $\mathfrak h$ will be denoted by $H_{\mathbb R}$; this
Cartan subgroup has
exactly $2^l$ connected components and the component of the identity is
denoted by $H=\exp({\mathfrak h})$. We let $\chi_i:=\chi_{\alpha_i} $
denote the roots characters defined on $H_{\mathbb R}$.
\end{Notation}
\begin{Example}
\label{tildeSL}

If $G=Ad(SL(n,\mathbb R) )$, then ${\tilde G}$ is isomorphic to
$SL(n,\mathbb R)$ for
$n$ odd and to $Ad(SL(n,\mathbb R)^{\pm}) $ for $n$ even. This example is
the underlying Lie group for the Toda lattices as shown in
\cite{casian:99}.
\end{Example}

\begin{Definition} The negative walls: Recall that  the compactified
isospectral manifold ${\tilde Z}(\gamma)_{\mathbb R} $ of the Toda lattice
is described in \cite{casian:99}
as a closure in $G/B^+$ of a generic $H_{\mathbb R}$ orbit.  Hence there is
an embedding
$f:H_{\mathbb R} \to {\tilde Z}(\gamma)_{\mathbb R} \subset G/B^+$. 

The exponential map $\exp: \mathfrak h \to H$ separates $H$, and consequently every
connected component of $H_{\mathbb R}$, into chambers.   If $\chi_i$ is a simple
root characters relative to a fixed dominant chamber then
 $\chi_{\alpha_i}$
can be extends to an adjacent chamber by
$s_{\alpha_j} \chi_{\alpha_i}=\chi_{\alpha_i}\chi_{\alpha_j}^{-C_{i,j} }$. This
defines a single function $\chi^*_i$ on an open dense subset of $H_{\mathbb R}$
which equals $\chi_{w(\alpha_i)}$ on each $w-$chamber for $w\in W$ (denoted
by $\phi_{w,i}$ in Definition 5.4 of  \cite {casian:99} ). 
The functions $|\chi_i^*|$ are well defined and continuous throughout 
$H_{\mathbb R}$ and the $\chi_i^*$ are well defined and  continuous
at all the  $\alpha_i$ walls and some of the $\alpha_j$ walls.
For example, if $\sigma $ is a permutation, then the corrersponding chamber 
in $SL(3,{\mathbb R })$ looks
like $ \{ (r_{\sigma (1)}, r_{\sigma (2)}, r_{\sigma (3)} ): |r_1| > |r_2 |> |r_3|  \}$
and $\chi^*_1= r_{\sigma (1) } r_{\sigma (2)}^{-1}$, 
$\chi^*_2= r_{\sigma (2) } r_{\sigma (3)}^{-1}$.

The
functions $\chi_i^*+1=0$ on $H_{\mathbb R}$ then determine a topological
subspace of ${\tilde Z}(\gamma)_{\mathbb R}$  whose closure we denote
$\tilde \Theta_{\{ i \} }$.  Similarly a subset $J\subset \Pi$
determines a topological space $\tilde \Theta_{J }$ by equations $\chi^*_i
+1=0$ for $\alpha_i \in J$. We call $\tilde\Theta_J$ the {\it negative wall}
associated with the set $J$
(see Subsection \ref{negativewalldef} for another definition in the
language of \cite{twisted} which does
not explicitly involve the Cartan subgroup).
\end{Definition}

\begin{conjecture} \label{conjecture1} There is a surjective continuous map
$f:\Theta_J
\to \tilde \Theta^a_J$. This map is a
 homeomorphism in an open dense
subset of $\Theta_J $. Whenever $\tilde \Theta^a_J$ happens to be
homeomorphic to a non-singular manifold
then $f$ is a homeomorphism.
\end{conjecture}

\begin{Example} In the case of ${\mathfrak sl}(3)$ all $\Theta_{\{ i \}}$
and $\tilde{\Theta}^a_{\{ i \}}$ are homeomorphic. They are both
homeomorphic to a circle (see Example \ref{A2exx} below).
For ${\mathfrak sl}(4)$ again $\Theta_{\{ i \} }$
is homeomorphic to $\tilde {\Theta}^a_{\{ i \} }$ for $i=1,3$ (details
in Example \ref{ex1} and Proposition \ref{A3divisors}).
However $\Theta_{\{ 2 \} }$
and $\tilde {\Theta}_{ \{ 2 \} }^a$ are not homeomorphic.   The situation
is described in Example \ref{ex2} together with Proposition \ref{A3divisors}
and  is
as follows. It is possible to desingularize
$\Theta_{ \{ 2 \} }$ so that the compact connected surface $\hat {\Theta}_{
\{ 2 \}}$ which is obtained is  non-orientable with Euler characteristic
$-2$.
Then there are maps $\hat {\Theta}_{\{ 2 \} } \to \Theta_{ \{ 2 \} } \to
\tilde {\Theta}_{\{2 \} }$ and $\hat {\Theta}_{ \{ 2 \} }$ now
 resolves the singularities of both
$\Theta_{ \{ 2 \} }$ and $\tilde \Theta_{ \{ 2 \} }$. Conjecture
\ref{conjecture1} needs to be sharpened  by
modifying $\tilde \Theta^a_J$ slightly so that
one always has homeomorphisms.  Below we propose such a modification for the
case when $J$ consists of one simple root.

It is now easy to see that $\Theta_{ \{ 1 \} }$ and $\tilde \Theta^a_{ \{ 1
\} }$ agree in the case of $A_3$. Figure \ref{A3:fig} shows the
eight polytopes $\Gamma_\epsilon$ corresponding to $2^l$ connected
components of $H_{\mathbb R}$. In fact
what is shown is the boundary of each polytope and the intersection of
$\Theta_{\{ i \} }$ for $i=1,2$.
However, the negative walls are also depicted by the same picture. The only
modification consists in
drawing the dotted lines or the solid grey lines  through the center of the
hexagons.
The actual negative walls are obtained by joining the dotted line or solid
grey line to the center of the polytope
through straight lines generating cones.  Hence $\tilde \Theta^a_{ \{ 1 \}
}$ intersected with each polytope consists of a disk in the form
of a cone joining the center of the polytope with the path described on the
boundary of the polytope
by the solid grey line. Gluings are described in detail in Definition
\ref{constructionGlue}. What results is a smooth compact surface.
\end{Example}

In order to introduce modifications to $\tilde \Theta^a_J$ we need to
describe its structure in more detail.
We do this by using the description of a manifold $M$ given
in \cite{twisted} which is homeomorphic to
$\tilde Z(\gamma)_{\mathbb R}$.  We review the construction of $M$
 and then define new topological spaces $\Theta^a_J$ in the case when $J$
conists of one simple root.

\begin{Definition}
\label{actsigns}
Let ${\mathcal E}$ be
the set of signs
${\mathcal E}=\left\{ (\epsilon_1, \cdots, \epsilon_l):\epsilon_k\in\{\pm\}
\right\}$. Then we define an action of $W$ on $\mathcal E$
by setting $s_i\epsilon=\epsilon^{\prime}$ where
 \[
 \epsilon_j^{\prime}=\epsilon_j\epsilon_{i}^{C_{j,i}}, \]
 which can be deduced from the $W$-action on the root character $\chi_i$
with $\epsilon_i={\rm sign}(\chi_i)$.
 The fact that this defines
an action which corresponds to the action of $W$ on the set of
connected components of a split Cartan subgroup of the real semisimple Lie
group ${\tilde G}$ can be found in \cite{casian:99}.

For any  $S\subset \Pi$ we let ${\mathbb D}(S)$ denote the set of all Dynkin
diagrams that have the simple roots in  $S$
 marked by $+$ or $-$. We also define an action of the group $W_S$ on this set
by making $w\in W_S$ act on the signs associated to the simple roots in $S$
as prescribed above. For example
 ${\circ}_{-}-{\circ}_{+}-{\circ} \in {\mathbb D}(S)$ with $S=\{ \alpha_1,
\alpha_2 \}$ and $s_1 ({\circ}_{-}-{\circ}_{-}-{\circ})
={\circ}_{-}-{\circ}_{+}-{\circ}$.
\end{Definition}

We now obtain actions of $W_S$ on  $ {\mathcal E}\times W$ and on ${\mathbb
D}(S) \times W$ given by
$\sigma (\epsilon , w)= (\sigma \epsilon, w\sigma^{-1})$ and $\sigma (\delta
, w)= (\sigma \delta, w\sigma^{-1})$.
The orbits of the $W_S$ action on $(\epsilon, w)\in {\mathcal E}\times W$
and  $(\delta, w) \in {\mathbb D}(S) \times W$ are denoted by $[\epsilon, w]_S$
and $[\delta, w])_S$ respectively with the sub-index $S$ dropped
when the set $S$ is clear from the context.  These $W_S$ orbits in the case
of ${\mathbb D}(S) \times W$
are the {\em full set of colored Dynkin diagrams} introduced in section 4 of
\cite{casian:99}.
The orbits of $W_S$  on $ {\mathcal E}\times W$ with $S=\{ \alpha_i \}$ are
used below to parametrize the  walls $\chi^*_i\pm 1=0$ intersected
with a fixed polytope.
The  walls  $\chi^*_i\pm 1=0$ in ${\tilde Z}(\gamma)_{\mathbb R}$ can be
parametrized by ${\mathbb D}(S) \times W$
with $S=\{ \alpha_i \}$.

\subsection{Review of the description of $\tilde Z(\gamma)_{\mathbb R}$ in
terms of the polytopes $\Gamma_\epsilon$} \label{revtwist}
Here we discuss the detailed description of negative walls in the connection
to the Painlev\'e divisors $\Theta_J$.
Let us first summarize the construction of the isospectral manifold of the
Toda lattice given
in \cite{twisted}.  Starting with a polytope $\Gamma$, other polytopes
$\Gamma_\epsilon$ are
constructed where $\epsilon\in {\mathcal E}$. These polytopes then
form a compact smooth manifold when they are glued together through their
boundaries. We now review the details.

In terms of
the description given in \cite{casian:99}, each $\Gamma_\epsilon$ has
interior that can
be made to correspond to a connected component of a split Cartan subgroup of
the real semisimple split Lie
group $\tilde G$.  {\em Chambers} and {\em walls} then refer to the action
of $W$ on
a Cartan subgroup, and the
internal chamber walls in the polytopes $\Gamma_\epsilon$ correspond to
walls of
this action ($\chi_i^* \pm 1=0)$.  When $\chi^*_i=-1$ then the chamber at the
other side
of the wall need not be the one obtained by application of $s_{\alpha_i}=s_i$.

\begin{Definition}
\label{constructionWalls}
Consider $\Gamma$ a convex polytope consisting of the
convex hull of a $W$ orbit  of
a regular element $x_o$ in $\mathfrak h$. We first denote $C^{\prime}_e$  the
dominant chamber in $\mathfrak h$ intersected with $\Gamma$ and
$\overline{C^{\prime}}_e$ the corresponding closure, and also denote
$C_{w}^{\prime}=w(C_e^{\prime})$. We define $C_{w}=\left\{ w \right\}\times
C^{\prime}_{w}$ and its closure $\overline {C}_{w}= \left\{ w \right\}
\times \overline {C^{\prime}}_{w}$.
The $^{\prime,\cdots}$ will refer to subsets of
$\Gamma$, and we have the convention:
\[
\left\{\cdots  \right\}^{\cdots}=\left\{ w \right\}\times \left\{\cdots
\right\}^{\prime \cdots}
\]
in all our notation concerning walls.
For each simple
root $\alpha_i$ we may consider the corresponding
$\alpha_i$ (internal) chamber wall intersected with $\overline
{C^{\prime}}_{w}$. Denote
this set by ${[w]}^{\prime ,\alpha_i,IN}$. Each external wall of the convex
hull of $Wx_o$ is
parametrized by a simple roots $\alpha_i$. We denote an {\it external}
wall of $\Gamma$ by
${[w]}^{\prime, \alpha_i, OUT}$ if it intersects all the {\it internal}
chamber walls except for
${[w]}^{\prime, \alpha_i, IN}$.

For any $J\subset \Pi$ we define the subsets of $\overline {C^{\prime}}_{w}$
of dimension $|\Pi\setminus J|$,
\[
\left\{
\begin{array} {lll}
{[w]}^{\prime, J,\Theta} &= \displaystyle{\bigcap_{\alpha_i\in J} }
{[w]}^{\prime, \alpha_i, \Theta},
~~ &{\text {if } J\not=\emptyset} ~,\\
{[w]}^{\prime , J,\Theta} &= C_{w}^{\prime}~, ~~ &{\text {if } J=\emptyset}~,
\end{array}\right.
\]
where $\Theta$ is either $OUT$ or $IN$. Thus we have the decomposition,
\[
{\overline {C^{\prime}}}_{w}=\displaystyle{\bigcup_{J\subset \Pi \atop
\Theta\in\{OUT,IN\}} {[w]}^{ \prime, J, \Theta}}.
\]
\end{Definition}

\begin{Definition}
\label{constructionGlue}
We will need to use  the action of $W$ on the set of signs ${\mathcal E
}$ of Definition \ref{actsigns}.
We now define gluing maps between the chamber walls
denoted by $\{\epsilon\}\times [w]^{\cdots}=\{\epsilon\}\times
\{w\}\times {[w]}^{\prime \cdots}$ as follows:
For the internal walls, we define
\[
\begin{array} {llll}
g_{w,i,IN} : &\displaystyle{\left\{\epsilon \right\}\times
 {[w]}^{\alpha_i, IN}} &\longrightarrow & \left\{s_{\alpha_i}\epsilon
\right\}\times
 {[ w s_{\alpha_i}]}^{\alpha_i, IN}      \\
& ~~\displaystyle{(\epsilon ,w, x )} &\longmapsto &(s_{\alpha_i}\epsilon ,
ws_{\alpha_i} , x)
\end{array}
\]
where note $ws_{\alpha_i}w^{-1}x=x$.
For the external walls, we define
\[
\begin{array} {llll}
  g_{w,i,OUT}: &\displaystyle{\left\{\epsilon \right\}\times
{[w]}^{\alpha_i, OUT}} &
  \longrightarrow &\left\{\epsilon^{(i)} \right\}\times
{[w]}^{\alpha_i,OUT}      \\
& ~~\displaystyle{(\epsilon , w, x )} &\longmapsto &(\epsilon^{(i)} , w , x)
\end{array}
\]
where $ \epsilon^{(i)}=(\epsilon_1, \cdots ,-\epsilon_i, \cdots ,\epsilon_l)$.

We denote ${\tilde M}$ the disjoint union of all the chambers endowed with
different signs,

\[
\tilde {M }=\bigcup_{w\in W \atop
\epsilon \in {\mathcal E}} \{w\epsilon\}\times \overline {C}_{w^{-1}} .
\]
We also denote $M$ the topological space
obtained from the disjoint union
in $\tilde M$ by gluing along the internal and
external walls using the maps
$g_{w,i,IN}$ and $g_{w,i,OUT}$. There is then a map
\[
z:\tilde{M}\to M.
\]
\end{Definition}

\subsection{The negative walls}
\label{negativewalldef}
We now give a precise definition of the negative wall. Let us first define:
\begin{Definition}
\label{gammaEpsilon}
First denote
 \begin{equation}
\displaystyle{ \tilde{\Gamma}_{\epsilon}=\bigcup_{w\in W} \left\{
w\epsilon \right\}\times \overline {C}_{w^{-1}}}~.
\nonumber
\end{equation}
We now let  $\Gamma_{\epsilon}$ denote the image of
$\tilde{\Gamma}_{\epsilon}$ in $M$. Then
 after the identifications in $M$, this space becomes a copy of
$\Gamma$.
\end{Definition}

\begin{Notation}
\label{standardneg}
Set $\epsilon^{\prime}=w^{-1}\epsilon$ and recall the action of $W$ and its
subgroups on pairs $(\epsilon^\prime, w)$  where
$\epsilon\in {\mathcal E}$ and $w\in W$ (Definition \ref{actsigns}).
Note that an  $\alpha_i$ wall ${[w]}^{\alpha_i, IN} $  which is the
intersection of two closed chambers
$\{ \epsilon^\prime \} \times \overline {C}_w $ and $\{ s_i\epsilon^\prime
\} \times \overline {C}_{ws_i} $
can be simply parametrized by the coset of $w$
in $[w]\in W/<s_i>$. To keep track of signs we need to consider the two
pairs $(\epsilon^\prime , w)$ and
$(s_i\epsilon^{\prime}, ws_i)$. This constitutes the orbit of
$(\epsilon^\prime , w)$ under the action of $W_{\left\{ s_i \right\}}$.
We have already  denoted this orbit by $[\epsilon^\prime , w]$
in Definition \ref{actsigns} and now this orbit
$[w^{-1}\epsilon, w]$  will also be used as a parameter to denote
the corresponding internal wall in  $\Gamma_\epsilon$.
This wall is called {\em negative } for $\alpha_i$ if
$\epsilon^{\prime}=(\epsilon^{\prime}_1,\dots , \epsilon^{\prime}_l)$
has $\epsilon^{\prime}_i=-1$.

For a set $J \subset \Pi$ one can also consider the orbit of $W_J$
denoted $[\epsilon^\prime, w]_J$ which will now denote the intersection:

\[
[\epsilon^\prime , w]_J = z\left( \displaystyle{\bigcap_{\sigma \in W_J} }
\{ \sigma \epsilon^\prime \} \times \overline {C}_{w\sigma^{-1}} \right)
\]
This parametrizes an intersection
of several walls. We call this $J$  multi-wall intersection 
$J-${\em negative} or just negative if it
is such that  for all $\alpha_i\in J$ there is 
$\sigma\in W_J$ such that $(\sigma \epsilon^{\prime})_i=-1$  where
$(\epsilon^\prime , w)$ is a representative
where $w$ has minimal length in its coset in $W/W_J$.

For any  $\alpha_i \in \Pi$  we consider the set $R_{\epsilon , J}$ given by
\[
R_{\epsilon , J}= \left\{ [\epsilon^\prime, w] : [\epsilon^\prime, w] \text{ is
negative  }
 \right\}
\]
When $J=\left\{\alpha_i \right\}$ we will just write  $R_{\epsilon ,i}$
\end{Notation}

 Consider the subspace of $M$ given by
\[
\tilde {\Theta}_J^a= z \left(\bigcup_{\epsilon\in {\mathcal E}, w\in
R_{\epsilon , J}}  [w^{-1}\epsilon ,w]_J \right).
\]

\begin{Example} \label{ex1} We can now describe the topology of $\tilde
\Theta^a_{\{ 1 \}  }$.  Since
$\tilde \Theta^a_{\{ 1 \}  }$ is smooth it will suffice to compute its Euler
characteristic.
That  $\tilde \Theta^a_{\{ 1 \}  }$ is not orientable will follow.

Walls in a fixed $\Gamma_\epsilon$ are parametrized as in Notation
\ref{standardneg}.
If we want to parametrize walls independently of each separate polytope, we
consider colored Dynkin diagrams as in \cite{casian:99}.  Intersections of
walls are obtained by coloring more simple roots with  $-$s or
$+$s.  Thus we consider walls as parametrized by the full set of colored
Dynkin diagrams
(Definition \ref{actsigns}).

The negative walls in $\Theta_{\{ 1 \} }$ can then be listed:
$[{\circ}_{-}-{\circ}-{\circ}, e]$, $[{\circ}_{-}-{\circ}-{\circ},2]$,
$[{\circ}_{-}-{\circ}-{\circ},3]$
$[{\circ}_{-}-{\circ}-{\circ}, 23]$,$[ {\circ}_{-}-{\circ}-{\circ},12]$ ,
$[{\circ}_{-}-{\circ}-{\circ}, 32],$ $ [{\circ}_{-}-{\circ}-{\circ},312]$
$[{\circ}_{-}-{\circ}-{\circ},123]$, $[{\circ}_{-}-{\circ}-{\circ},232]$,
$[{\circ}_{-}-{\circ}-{\circ},1232]$, $[{\circ}_{-}-{\circ}-{\circ}, 2312]$,
$[{\circ}_{-}-{\circ}-{\circ},12132].$

Now boundaries must be considered.
For example the boundaries of $[{\circ}_{-}-{\circ}-{\circ}, e]$ are $[
{\circ}_{-}-{\circ}_{-}-{\circ}, e], [ {\circ}_{-}-{\circ}-{\circ}_{-}, e],
[{\circ}_{-}-{\circ}_{+}-{\circ},e], [{\circ}_{-}-{\circ}-{\circ}_{+}, e]$.
Therefore all these walls are part of $\Theta_{\{ 1 \} }$.
However $[{\circ}_{-}-{\circ}-{\circ} ,2]$  produces
$[{\circ}_{-}-{\circ}_{-}-{\circ}, 2]$. Since the Weyl group $W_S$ now
includes $s_2$ then we can write this wall as
$[{\circ}_{+}-{\circ}_{-}-{\circ}, e]$ because $s_2
({\circ}_{-}-{\circ}_{-}-{\circ})={\circ}_{+}-{\circ}_{-}-{\circ} $.
Therefore the wall $[{\circ}_{+}-{\circ}_{-}-{\circ}, e]$ must also be
included in $\tilde \Theta^a_{\{ 1 \}  }$.
Taking this into account we can easily count all the cells of $\tilde
\Theta^a_{\{ 1 \}  }$.
All the 1-cells of the form
$[{\circ}_{\epsilon_1}-{\circ}_{\epsilon_2}-{\circ}, w]$
except $\epsilon_1=\epsilon_2=+$ appear. This gives $3\times |W/W_{\{ s_1,
s_2 \} } |=3\times 4$
such cells. We also get all the cells of the form
$[{\circ}_{-}-{\circ}-{\circ}_{\pm}, w]$ , a total of $2\times |W/W_{\{ s_1,
s_3 \} } |
=2\times 6$. Hence a total of $24$ cells of dimension one. Finally there are
$7$ cells of dimension $0$.
The Euler characteristic obtained is $12-24+7=-5$. Since $\tilde
\Theta^a_{\{ 1 \} }$ is
homeomorphic to a smooth compact surface, this completely describes its
topology.

Note that the actual boundary maps involved in a homology computation are as
described in section 4 of \cite{casian:99}.
\end{Example}

\subsection{A graph associated to the negative walls in $\Gamma_\epsilon$}
Let us define a graph to describe the negative walls for a fixed
$\alpha_i\in \Pi$.
\begin{Definition} The graph ${\mathbb G}(\epsilon)$:
\label{deadend}

We  consider a graph ${\mathbb G}(\epsilon)$ having  vertices
$(\epsilon^\prime, w)$ with $\epsilon^\prime=w^{-1}\epsilon$.

\begin{itemize}
\item[(a)] If all $\epsilon_j=-$ then all the pairs $(\epsilon^\prime , w)$,
$(s_j\epsilon^\prime, ws_j)$ are edges.
\end{itemize}

We now describe the edges when not all $\epsilon_j$ are negative.

First for all semisimple Lie algebras of rank  $l\leq 3$:

$(\epsilon^\prime , w)$, $(s_j\epsilon^\prime, ws_j)$
is an edge if and only if one of the following is satisfied:

\begin{itemize}
\item[(b)]  $i\not=j$, $C_{j,i}\not=-2$ ,  $\epsilon^\prime_j
=+$
\item[(c)] $i\not=j$, $C_{j,i} =-2$, $\epsilon^\prime_j =-$
\item[(d)]   $i\not=j$, $C_{i,j}=0$, $\epsilon^\prime_i=-$
\end{itemize}

Note that if we fix $i$ and $j$ then the corresponding subdiagram of the
Dynkin diagram has rank two.
We will show below that these conditions lead to the correct description of
the divisors
$\Theta_i$ in the rank two cases.
The condition  d) of Definition \ref{deadend} will correspond to
the case of $A_1\times A_1$.
If we consider only Lie algebra of types $A$, $D$, $E$ and $G_2$,
the conditions simplify to:
\begin{itemize}
\label{deadendrank2}
\item[(a')] \label{deadone1} $\alpha_j \in \Pi(\epsilon)$
\item[(b')]\label{deadtwo1} $i\not=j$,   $\epsilon^\prime_j =+$
\item[(c')] \label{deadfour1} $i\not=j$, $s_j$ commutes with $s_i$ and
$\epsilon^\prime_i=-$
\end{itemize}
The case $\alpha_i=\alpha_2$ in $B_2$ is some kind of  exception which
requires a separate rule given in condition c).

In general,if the rank is $n$ given a subset $S\subset \Pi$ and $\epsilon\in {\mathcal E}$
we denote $\epsilon_S$ the restriction formed by  the ordered
$|S|$-tuple consisting only of the $\epsilon_k$ with $\alpha_k\in S$.

 We assume that all the edges of the graph have been defined
for rank $|S|< n$. The pair  $(\epsilon^\prime , w)$, $(s_j\epsilon^\prime,
ws_j)$
is an edge if there is $S \subset \Pi$ with $|S| < n$,  $s_i, s_j\in S$
and there is $\sigma \in W_S$ with $w=w_1\sigma$, $\ell (w_1)+\ell
(\sigma)=\ell(w)$
and $(\epsilon_S^\prime , \sigma)$, $(s_j\epsilon_S^\prime, \sigma s_j)$
form an edge
in the case of the split Lie subalgebra determined by $S$.

\end{Definition}

We now break up $R_{\epsilon , i}$ as a disjoint union of subsets consisting
of negative walls belonging to the
same connected component of the graph ${\mathbb G}(\epsilon)$. We thus
obtain a set $I(\epsilon)$ consisting
of subsets of $R_{\epsilon , i}$. The disjoint union $\bigcup_{\alpha \in
I(\epsilon) } \alpha $ equals $R_{\epsilon , i}$.

\begin{Definition} The graph ${\mathbb G}$:
We now define a graph whose vertices are the elements $\alpha\in
I(\epsilon)$ for $\epsilon\in {\mathcal E}$.
If $\alpha \in  I(\epsilon_1)$ and $\beta\in I(\epsilon_2)$ , then there is
an edge joining $\alpha$ to $\beta$ if and only if there is  $w$  such that

\begin{itemize}
\item[(i)] $[w^{-1}\epsilon_1, w] \in \alpha $ , $[w^{-1}\epsilon_2, w] \in
\beta $
\item[(ii)] Denoting $w^{-1}\epsilon_1=\epsilon^\prime$ then we have:
$w^{-1}\epsilon_2=({\epsilon^\prime} )^{(i)}$
\end{itemize}
\end{Definition}

\begin{Example} \label{A2exx} Consider the case of $A_2$ and $J=\{ \alpha_1
\}$.
If   $\epsilon=(--)$ condition a) in Definition \ref{deadend} applies;
however, as it turns out, condition
b) alone will suffice in this case. We have the
following edges indicated by $\to$ connecting the only two negative walls.
$((--),e) \to ((-+), s_1)\to ((-+), s_1s_2)$. We have the following set of
negative
walls  $R_{(--), 1}= \{ [(--), e] , [(-+), s_1s_2] \}.$ We obtain that
 $I{(--)}$ consists of one single element $\alpha^{--} =\{ [(--), e] ,
[(-+), s_1s_2] \}$.

For $\epsilon=(-+)$ we obtain $R_{(-+), 1}=\{ [(-+), e], [(-+), s_2] \} $
and $I{(-+)}$ consists
of one single element $\alpha^{-+} =\{ [(-+), e], [(-+), s_2] \} $

For  $\epsilon=(+-)$ we have $R_{(+-), 1}=\{ [(--),s_2] \to [(--),
s_1s_2] \}$ and again there is one single
element $\alpha^{+-}.$

The graph $\mathbb G $ consists of a \lq\lq cycle\rq\rq  $\alpha^{--} \to
\alpha^{-+} \to \alpha^{+-} \to  \alpha^{--}$.
For example, there is an edge $\alpha^{--} \to \alpha^{-+}$  because
$[(-+), e] \in \alpha^{--}\cap \alpha^{-+}$.
If one consider the topological space consisting of the corresponding walls
then what results is a circle in agreement
with what was found in Propopsition \ref{A2topology}. This corresponds to
Figure \ref{A2:fig} where the divisor indicated by the
number $1$ is replaced with two walls -straight lines- joining at the center
of the hexagons.  The edges of
$\mathbb G$ correspond to intersections with the boundaries of the hexagons
$\Gamma_\epsilon$, that is with \lq\lq subsystems\rq\rq.
\end{Example}

\begin{Example}
Consider the case of $G_2$  and $J=\left\{ s_1 \right\}$ with
$\epsilon=(+-)$.  The
negative walls are parametrized by
$$[(--), s_2],[ (-+), s_2s_1s_2], [(--), s_1,s_2], [(-+), s_1s_2s_1s_2].$$

Note that   $[(--), s_2],[ (-+), s_2s_1s_2]$ are in the same connected
component of ${\mathbb G}(\epsilon)$
since $$(--, s_2)\to (-+, s_2s_1)\to (-+, s_2s_1s_2);$$ where $\to$
indicates an edge. However this process reaches
a dead-end when we apply $s_1$
once more since one obtains $(--, s_2s_1s_2s_1)$ but $s_2$ cannot be applied
at this point because $\epsilon_2=-1$.  The connected
component of the graph ${\mathbb G}(+-)$ which contains $[(--), s_2]$ then
consists of
$$(--, s_2)\to (-+, s_2s_1)\to (-+, s_2s_1s_2) \to (--, s_2s_1s_2s_1).$$
From here $$\alpha^{+-}=\left\{ [(--), s_2],[ (-+), s_2s_1s_2] \right\}$$
and similarly there is another set of negative walls $$\beta^{+-}=\left\{
[(--), s_1s_2],[(-+), s_1s_2s_1s_2] \right\}.$$
We have $I(+-)=\left\{\alpha^{+-}, \beta^{+-} \right\}$.

 For $\epsilon=(-+)$, $\Pi(\epsilon)=\left\{s_1 \right\}$ one obtains
$\alpha^{-+}=\left\{ [(-+),e], [(-+), s_2] \right\}$,
 $\beta^{+-}=\left\{ [(--),s_1s_2s_1s_2 ], [(--), s_2s_1s_2s_1s_2]
\right\}$.  For $\epsilon=(--)$ all the negative walls are in a single
connected
 component. However, here, unlike what happens in the $A_2$ example one
requires  condition a) of Definition \ref{deadend}
 with $\Pi(--)=\{ s_1, s_2 \}$. This allows the application of $s_2$
independently of the sign $\epsilon_2^\prime$. We have
 $\alpha^{--}=\left\{ [(--), e] , [(-+), s_1,s_2], [(--), s_2s_1s_2], [(-+),
s_2s_1s_2s_1s_2] \right\}$ and $I(--)=\left\{\alpha^{--} \right\}$.

 The graph ${\mathbb G}$ has nodes given by $\left\{ \alpha^{+-}.
\beta^{+-}, \alpha^{-+}, \beta^{-+}, \alpha^{--}  \right\}$.  The edges
 are $\alpha^{+-} \to \alpha^{-+}$, $\beta^{+-} \to \beta^{-+}$ and
$\alpha^{--} \to x$ for $x=\alpha^{+-}, \beta^{+-}, \alpha^{-+},
\beta^{-+}$.
 This gives a total of six edges.

 When one considers the topology of the sets of walls involved and the edges
are regarded as the only gluings: $\alpha^{--} $ consists
 of two intersecting line segments and all the others consist of segments.
What then results is a figure $8$. The 6 edges are
 the intersection of this figure $8$ with the boundaries of the polytopes
$\Gamma_\epsilon$ (subsystems).
 Note that  segments  associated to $\alpha^{+-}$ and $\beta^{+-}$ are being
regarded as {\em disjoint}.
 However the two segments forming $\alpha^{--}$ are not disjoint  because
they form part of one single connected component  of ${\mathbb G}(--)$.
 The topological space associated to these graphs and the negative walls
will be made precise below for a general semisimple
 Lie algebra.
\end{Example}

\begin{Example}
We now consider the case of $B_2$, $J=\left\{ s_2
\right\}$ and $\epsilon= (+,-)$.
We have edges $(+-,e)\to (+-,s_1) \to (+-,s_1s_2)$  but  $(+-,s_1s_2)$ is a
dead-end because $s_1$ cannot be applied since
$\epsilon_1=+$ and $C_{1,2}=-2$.  We also have an edge $(+-,e)\to (+-,s_2)$
which leads to a dead-end for the same reason. This
gives a set $\alpha^{+-}=\left\{ [(+-),e] ,[(+-),s_1] \right\}$. Another
connected component of the graph produces
$\beta^{+-}=\left\{ [(+-), s_2s_1] , [(+-), s_1s_2s_1] \right\}$.

For $\epsilon=(--)$ one obtains $\alpha^{--}=\left\{ [(--), e] , [(--),
s_1s_2s_1]\right\}$  and for $\epsilon=(-+)$
$\alpha^{-+}=\left\{ [(-+), s_1] , [(-+), s_2s_1] \right\} $. Hence a graph
results with edges $\alpha^{--}\to x$
 and $\alpha^{-+}\to x$  where $x=\alpha^{+-}, \beta^{+-}$ giving a
total of four edges in ${\mathbb G}$.
 Again we consider the topology of the sets of walls involved and, as in
 the previous examples,  the edges in ${\mathbb G}$
are regarded as the only gluings between
 these segments. We obtain four segments corresponding to
 the elements in $I(\epsilon)$  giving rise to a circle that
 intersects the boundaries of the $\Gamma_\epsilon$ at four points (the
edges of the graph ${\mathbb G}$). This corresponds to $\Theta_{\{ 1 \} }$ in
Proposition \ref{C2topology}.

\end{Example}

\subsection{The spaces of negative walls $\Theta^a_i$ }
Fix an element $\alpha\in I(\epsilon)$. We consider  the disjoint union
\[
\bigcup_{\alpha\in I(\epsilon),[\epsilon^\prime , w]\in \alpha , \epsilon\in
{\mathcal E} }\left\{\alpha \right\}\times [w^{-1}\epsilon , w].
\]
We define gluings for any pair $\alpha,\beta$ which are joined by an edge of
the graph ${\mathbb G}$.
\[
\begin{array} {llll}
{ g_{}}: & {(\alpha, [\epsilon^\prime ,w])} & \longrightarrow & [(\beta,
{\epsilon^\prime}^{(i)} w\sigma^{-1} ])  \\
 {} & {(\alpha, \epsilon^\prime ,w, x )}  & \longmapsto & ({\beta,
\epsilon^\prime}^{(i)} , w , x) \\
\end{array}
\]

\begin{conjecture} \label{conjecture2} There is a homeomorphism $g:\Theta_i
\to \Theta_i^a$.
\end{conjecture}

\begin{Example}
 \label{ex2}  The topology of  $\Theta^a_{\{ 2 \}  }$ in the case
of $A_3$  can  be computed
explicitly and shown to
correspond to $\Theta_{\{ 2 \} }$. We first compute
the Euler characteristic of $\tilde \Theta^a_{\{ 2 \}  }$  using
the method in Example \ref{ex1}. One obtains twelve 2-cells, twenty four 2
cells and seven 1-cells
giving the same Euler characteristic as in the case of $\tilde \Theta^a_{\{
1 \}  }$.  However the sets
$I(\epsilon)$ for $\epsilon=(+-+)$ and $\epsilon=(-+-)$ contain two
elements. This can be seen in Figure \ref{A3:fig} where
the corresponding paths of dotted lines are disconnected. The recipe for the
construction of $\Theta^a_{\{ 2 \}  }$
corresponds to separating the two cones obtained by joining these paths to
the center of each of these polytopes.
This introduces two additional points! Hence the Euler characteristic for
$\Theta^a_{\{ 2 \}  }$ becomes -3.

One now notes that $\Theta^a_{\{ 2 \}  }$ remains singular as can be seen
in Figure \ref{A3:fig}
where in the boundary of the polytope $\Gamma_{---}$ there
are two disconnected paths of dotted lines. It is possible to resolve this
singularity by simply
separating the center of this polytope into two separate points. This gives
rise to a
compact surface of Euler characteristic -2 since one additional point is
added. The resulting
surface can be seen to be non-orientable.  We thus obtain that $\Theta^a_{\{
2 \}  }$  is homeomorphic to
$\Theta_{\{ 2 \}  }$.  The compactification of the isospectral manifold
of $A_2$  reappears but only as a resolution of singularities of the
Painlev\'e divisor.
\end{Example}

\bibliographystyle{amsalpha}

\end{document}